\documentclass[useAMS,usenatbib]{mn2e}
\usepackage{aas_macros}
\usepackage{amsmath,amssymb}
\usepackage{bm}
\usepackage{color}
\usepackage[dvipdfmx]{hyperref}
\usepackage{graphicx}
\usepackage{enumerate}
\usepackage{ulem}
\newcommand{\nnmb}{\nonumber\\}

\newcommand{\ds}{\displaystyle}
\newcommand{\mr}[1]{\mathrm{#1}}
\newcommand{\cmr}[1]{\,\mathrm{#1}}

\newcommand{\Hii}{H\,{\sc ii} }

\usepackage{amsmath}	

\begin{document}
\author[K. Sugimura et al.
]{Kazuyuki Sugimura,$^1$\thanks{E-mail: sugimura@astr.tohoku.ac.jp} 
Takashi Hosokawa,$^{2,3,4}$
Hidenobu Yajima$^{1,5}$ \newauthor
and Kazuyuki Omukai$^{1,3}$
\\
$^1$Astronomical Institute, Tohoku
University, Aoba, Sendai 980-8578, Japan \\ 
$^2$Department of Physics, Kyoto University, Sakyo, Kyoto 606-8502, Japan\\
$^3$Kavli Institute for Theoretical Physics, University of California, Santa Barbara, California 93106, USA\\
$^4$Department of Physics and Research Center for the Early Universe, the University of Tokyo, Bunkyo, Tokyo 113-0033, Japan\\
$^5$Frontier Research Institute for Interdisciplinary Sciences, Tohoku University, Aoba, Sendai 980-8578, Japan
} 
\title[BH growth under anisotropic radiation]
{Rapid Black Hole Growth under Anisotropic Radiation Feedback}
\maketitle
\topmargin-1cm

\begin{abstract}
Discovery of high-redshift ($z > 6$) supermassive black holes (BHs) may
indicate that the rapid (or super-Eddington) gas accretion has aided
their quick growth. Here, we study such rapid accretion of the
primordial gas on to intermediate-mass ($10^2 - 10^5~M_\odot$) BHs under
anisotropic radiation feedback.  We perform two-dimensional radiation
hydrodynamics simulations that solve the flow structure across the Bondi
radius, from far outside of the Bondi radius down to a central part
which is larger than a circum-BH accretion disc.  The radiation from the
unresolved circum-BH disc is analytically modeled considering
self-shadowing effect.  We show that the flow settles into a steady
state, where the flow structure consists of two distinct parts: (1)
bipolar ionized outflowing regions, where the gas is pushed outward by
thermal gas pressure and super-Eddington radiation pressure, and (2) an
equatorial neutral inflowing region, where the gas falls toward the
central BH without affected by radiation feedback.  The resulting
accretion rate is much higher than that in the case of isotropic
radiation, far exceeding the Eddington-limited rate to reach a value
slightly lower than the Bondi one.  The opening angle of the equatorial
inflowing region is determined by the luminosity and directional
dependence of the central radiation.  We find that photoevaporation from
its surfaces set the critical opening angle of about ten degrees below
which the accretion to the BH is quenched.  We suggest that the
shadowing effect allows even stellar-remnant BHs to grow rapidly enough
to become high-redshift supermassive BHs.
\end{abstract}

\section{Introduction}
\label{sec:intro}

Discovery of high-$z$ ($z\gtrsim6$) quasars suggests that supermassive
black holes (SMBHs) already exist when the age the Universe is less than
$1\cmr{Gyr}$ \citep[see,
e.g.,][]{Fan:2001aa,Willott:2010aa,Mortlock:2011aa,Venemans:2013aa,Wu:2015aa}.
This poses a question about the formation mechanism of SMBHs in such a
short interval. Among the scenarios for the SMBH seed formation
\citep[see, e.g.,][for a review]{Volonteri:2012ab,Haiman:2013aa},
including the dense stellar cluster scenario \citep[see, e.g., ][and
reference
therein]{Omukai:2008aa,Devecchi:2009aa,Katz:2015ab,Yajima:2016aa},
following two are the most studied: the direct collapse BH (DCBH) and
the population III (Pop III) remnant BH scenarios.

In the former scenario, supermassive stars of $\sim 10^5\,M_\odot$
collapse to form seed BHs with approximately the same mass.
Specifically, supermassive stars are envisaged to form in exceptional
environments in the high-$z$ Universe, for example, in atomic-cooling
halos where the $\mr{H_2}$ cooling is totally suppressed by very strong
far ultraviolet (FUV) irradiation \citep[e.g.,][]{Sugimura:2014aa}.
While the seed BHs in this case are rather massive with $\sim
10^5\,M_\odot$, their number density might be too small to explain all
the observed high-$z$ SMBHs due to the stringent necessary conditions
\citep{Dijkstra:2008aa,Dijkstra:2014aa,Agarwal:2012aa,Sugimura:2014aa,Sugimura:2016aa,Inayoshi:2015ab,Chon:2016aa}.

In the latter scenario, the remnant BHs of Pop III stars
\citep{Yoshida:2008aa, Hosokawa:2011aa,Hosokawa:2016aa} are thought as
SMBH seeds \citep{Alvarez:2009aa,Jeon:2012aa}.  Contrary to the DCBH
scenario, they are abundant but the problem is whether they can actually
grow to the SMBHs from smaller initial mass of $\lesssim 10^3\,M_\odot$
\citep{Susa:2014aa,Hirano:2015aa} within the available time.  Although
BHs can acquire the mass by collisions with other BHs
\citep{Tanikawa:2011aa}, the BH collisions often result in ejection of
the merged BHs from the host halo due to the recoil of gravitational
wave emission \citep[e.g.,][]{Baker:2006aa,Koppitz:2007aa}.  Thus, the
feasibility of this scenario relies on whether the rapid accretion on to
seed BHs is possible or not
\citep{Madau:2014aa,Alexander:2014aa,Volonteri:2015aa}.

Recently, a number of authors have studied the BH accretion under
radiation feedback
\citep[e.g.,][]{Milosavljevic:2009aa,Milosavljevic:2009ab,
Park:2011aa,Park:2012aa,Park:2013aa}.  They solve the gas dynamics over
the scale of the Bondi radius, where the accretion rate on to the
circum-BH disc is physically determined.  Although the central circum-BH
disc is not spatially resolved, subgrid models that provide analytic
prescriptions of its emissivity have been used.  They have shown that
the accretion rate is significantly reduced to $\lesssim1\%$ of that
without radiation feedback (i.e., the Bondi rate) in case with modest BH
mass and ambient density (e.g., $10^2\,M_\odot$ and
$10^5\cmr{cm^{-3}}$). Only in case with very high BH mass and/or ambient
density (e.g., $10^4\,M_\odot$ and $10^5\cmr{cm^{-3}}$), the accretion
rate reaches to the Bondi value because of inefficient radiation
feedback, as recently shown by \cite{Inayoshi:2016ac} \citep[see
also][for other mechanisms of efficient
accretion]{Li:2011aa,Pacucci:2015ab,Park:2016ab}.  However, all those
calculations assume isotropic radiation (in either one- or
two-dimensional simulations), whereas in reality the radiation from the
BH accretion disc should be anisotropic. The flow structure will be
significantly altered in such anisotropic radiation field.  Although the
BH accretion under anisotropic radiation has been studied in the context
of active galactic nuclei (AGN) with the BH mass $\ga 10^6\,M_\odot$
\citep{Proga:2007aa,Kurosawa:2009aa,Novak:2011aa,Barai:2012aa}, the
nature of accretion on to stellar-mass BHs would be quite different.

The anisotropic BH irradiation has been examined with different models
of the BH accretion discs, including the ``standard disc'' for moderate
accretion rates \citep{Shakura:1973aa}, and ``slim disc'' for the higher
rates \citep{Abramowicz:1988aa}. In particular, recent multi-dimensional
simulations have investigated inner structure of the slim disc within
roughly a hundred Schwarzschild radii, showing that the accretion rates
can indeed exceed the Eddington-limited rate
\citep[e.g.,][]{Ohsuga:2005aa,Jiang:2014aa,McKinney:2014aa,
Fragile:2014aa,Takahashi:2015aa,Sc-adowski:2016aa}.  These studies show
that the high-energy photons are predominantly emitted in polar
directions from the inner part of the disc.  However, the outer
structure of the disc, which is not solved in the above simulations,
should also modify the anisotropic radiation field. For instance, disc
winds such as the line-driven AGN winds launched from the outer region
will absorb a part of photons coming from the inner region
\citep[e.g.,][]{Proga:2000aa,Proga:2004aa,Nomura:2016aa}.  Since
numerical simulations solving the whole structure of the disc are still
infeasible, it is very uncertain how much anisotropy the BH accretion
discs actually create.

In this paper, we will investigate accretion of the primordial gas on to
BHs under the anisotropic radiation feedback from the central circum-BH
accretion discs, considering the shadowing effect by the outer part of
the discs. We perform a set of proof-of-concept two-dimensional (2D)
radiation hydrodynamics (RHD) simulations, assuming that BHs are
initially embedded in homogeneous and static media.  We do not attempt
to simulate the realistic directional dependence of BH irradiation in
consideration of its high uncertainties; instead, we model it in a
simple fashion to study how the anisotropy of radiation changes the
nature of accretion flows.  As confirmed later by our results, the
shadowing effect dramatically enhances the accretion rate.  This
mechanism might give a new pathway from the remnant BHs of Pop III stars
to SMBHs within a limited timescale of $\la 1\cmr{Gyr}$ after the Big
Bang.
 
The paper is organized as follows. In Sec.~\ref{sec:sph_acc}, we briefly
review the basics of spherical gas accretion on to a BH.  In
Sec.~\ref{sec:num-method}, we describe the numerical method and cases
considered.  In Sec.~\ref{sec:result}, we present the main results of
our simulations.  The conclusions and discussions are given in
Sec.~\ref{sec:conclusion}.

\section{Basics}
\label{sec:sph_acc}

For later reference, we first briefly summarize the basics of
spherical gas accretion on to a central BH under radiation
feedback. 
We consider a system where a BH is embedded in a static and homogeneous medium.
We take the BH mass $M_\mr{BH}=10^3\,M_\odot$, ambient density
$n_\infty=10^5\,\mr{cm^{-3}}$ and ambient temperature
$T_\mr{HI}=10^4\,\mr{K}$
as a fiducial parameter set.

If we ignore the effect of feedback, the mass accretion will proceed at
the Bondi rate in this case,
\begin{align}
\dot{M}_\mr{B}
&=\frac{4\pi\lambda_\mr{B}\rho_\infty G^2 M_\mr{BH}^2}{c_\mr{s,HI}^3}\nnmb
&=1.7\times 10^{-3}
\,\left(\frac{n_\infty}{10^5\,\mr{cm^{-3}}}\right)\nnmb
&\qquad \times \left(\frac{M_\mr{BH}}{10^3\,M_\odot}\right)^{2}
 \left(\frac{T_\mr{HI}}{10^4\,\mr{K}}\right)^{-3/2}
 \mr{M_\odot\, yr^{-1}}
 \,,\label{eq:1}
\end{align}
where we take
$\lambda_\mr{B}=(1/4)\left[2/(5-3\gamma)\right]^{(5-3\gamma)/2(\gamma-1)}
= 1.12$ assuming the gas is isothermal (the polytropic index
$\gamma=1$).  For a neutral primordial gas with helium-to-hydrogen ratio
in the number of nuclei $y_\mr{He}=0.0972$, the mean molecular weight
$\mu=(1+4y_\mr{He})/(1+y_\mr{He})=1.3$, the mass density of the medium
$\rho_\infty=n_\infty(1+4y_\mr{He})m_\mr{p}=2.3\times10^{-19}\cmr{g\,cm^{-3}}$
with $m_\mr{p}$ the proton mass and the (isothermal) sound speed
$c_\mr{s,HI}=(k_\mr{B}T_\mr{HI}/\mu m_\mr{p})^{1/2}
=8.1\,\left(T_\mr{HI}/10^4\,\mr{K}\right)^{1/2} \,\mr{km\, s^{-1}}$.
The Bondi radius, defined as
\begin{align}
 r_\mr{B}
&=\frac{GM_\mr{BH}}{c_\mr{s,HI}^2}\nnmb
&=1.4\times 10^{4} 
 \left(\frac{M_\mr{BH}}{10^3\,M_\odot}\right)
 \left(\frac{T_\mr{HI}}{10^4\,\mr{K}}\right)^{-1}
 \mr{AU}\,,
\label{eq:2}
\end{align}
demarcates the inner region where the gravitational energy dominates the
thermal energy and the outer region where the thermal energy dominates.
Correspondingly, the gas is approximately in free fall inside
$r_\mr{B}$, whereas the pressure equilibrium is almost achieved outside.

The Eddington luminosity $L_\mr{E}$ is the critical luminosity above
which the outward radiation force via the Thomson scattering exceeds the
inward gravitational pull of the BH in fully ionized hydrogen gas,
\begin{align}
L_\mr{E}=\frac{4\pi G M_\mr{BH} c m_\mr{p}}{\sigma_\mr{T}}
&=3.3\times 10^7
\left(\frac{M_\mr{BH}}{10^3M_\odot}\right) L_\odot\,,
\label{eq:12}
\end{align}
where $\sigma_{\mr{T}}$ is the Thomson scattering cross section.  Note
that the Eddington luminosity does not always provide physical limit
because gas pressure is not considered in the above argument.  In
addition, the radiation force becomes less effective in a partially
ionized gas.

The (efficiency-independent) Eddington-limited accretion rate is defined
as
\begin{align}
 \dot{M}_\mr{E}
&=\frac{L_\mr{E}}{c^2}
=2.2\times 10^{-6} \left(\frac{M_\mr{BH}}{10^3M_\odot}\right)\,M_\odot\, \mr{yr^{-1}}\,,
\label{eq:8}
\end{align}
and the condition for the luminosity to be sub-critical can be rewritten
as $\dot{M}<\dot{M}_\mr{E}/\eta$ with the radiative efficiency
$\eta$. The radiative efficiency $\eta\approx 0.1$ for a standard
accretion disc is widely used in the previous works. \citep[see,
e.g.,][]{Milosavljevic:2009aa,Park:2011aa,Park:2012aa}.  Note that in
some literatures the efficiency-dependent Eddington-limited accretion
rate, $\dot{M}_\mr{E}/\eta$ in our definition, is used instead.  For
large $M_\mr{BH}$ and/or $n_\infty$, the Bondi rate $\dot{M}_\mr{B}$ can
be much larger than the Eddington rate $\dot{M}_\mr{E}$, e.g.,
$\dot{M}_\mr{B}/\dot{M}_\mr{E}=7.8\times 10^{2}
\,\left(n_\infty/10^5\,\mr{cm^{-3}}\right)
\left(M_\mr{BH}/10^3\,M_\odot\right)$, because $\dot{M}_\mr{B}$ is
proportional to $M_\mr{BH}^2\,n_\infty$ while $\dot{M}_\mr{E}$ to
$M_\mr{BH}$.

High energy photons emitted by the BH accretion disk create a
surrounding \Hii bubble.  With the power-law spectrum $L_\nu \propto
\nu^{-1.5}$, which is often postulated in the literature, the
photoionized gas is heated up to $T_\mr{HII}\sim 7\times10^4\cmr{K}$
owing to helium ionization heating.  The high thermal pressure of the
\Hii bubble, together with the outward radiation pressure, can
significantly reduce the accretion rate
\citep{Milosavljevic:2009aa,Park:2011aa,Park:2012aa}.  The size of the
\Hii bubble is estimated by the Str\"omgren radius,
\begin{align}
  r_\mr{HII} &=
6.8\times 10^{4}
\left(\frac{T_\mr{HII}}{7\times10^4\cmr{K}}\right)^{\frac{1}{3}}\nnmb
&\qquad\times\left(\frac{L}{3.3\times10^7L_\odot}\right)^{\frac{1}{3}}
\left(\frac{n_\mr{HII}}{10^5\,\mr{cm^{-3}}}\right)^{-\frac{2}{3}} 
 \cmr{AU}\,,
\label{eq:13}
\end{align}
which is obtained by equating the ionizing photon emissivity
$\dot{N}_\mr{ion}=\int_{\nu_\mr{T}}^\infty d\nu L_\nu/h\nu$
($=L/3h\nu_\mr{T}$ for the above spectrum with $L_\nu \propto
\nu^{-1.5}$) with the recombination rate within the \Hii bubble
$\alpha_\mr{B}(4\pi/3)r_\mr{HII}^3 n_\mr{HII}^2$, where $n_\mr{HII}$ is
the number density of hydrogen nuclei inside the bubble,
$h\nu_\mr{T}=13.6\cmr{eV}$ the hydrogen ionization energy, and
$\alpha_\mr{B}=4.6\times10^{-14}\cmr{cm^3\,s^{-1}}$ the case B hydrogen
recombination coefficient at $7\times10^4\cmr{K}$
\citep{Ferland:1992aa}.  In the above, we take
$n_\mr{HII}=n_\infty\,(=10^5\cmr{cm^{-3}})$ and $L=L_\mr{E}$ as
reference values.

Suppose that an ionizing source is suddenly turned on at the centre.
The ionization front first propagates up to $r_\mr{HII}$ with
$n_\mr{HII}$ kept almost constant. Then the bubble expands until
pressure equilibrium with the surrounding medium is reached with
$n_\mr{HII}=(c_\mr{s,HI}/c_\mr{s,HII})^2\,n_\infty \sim 0.07\,n_\infty$,
where $c_\mr{s,HII} = (2T_\mr{HII}/T_\mr{HI})^{1/2}c_\mr{s,HI}$ is the
sound speed of the ionized gas, with the factor of $2$ accounting for
the increase of the particle number by ionization.  Note that we have
neglected the effect of helium in estimating $r_\mr{HII}$ and
$n_\mr{HII}$, because it modifies them only slightly.  For the flow from
the \Hii bubble in pressure equilibrium with the surrounding neutral
medium, the Bondi radius and rate are given by $r_\mr{B,HII}\sim
0.07\,r_\mr{B}$ and $\dot{M}_\mr{B,HII} \sim 1\times 10^{-3}
\dot{M}_\mr{B}$, respectively. This clearly shows that the
photo-ionization feedback can considerably suppress the accretion.

In order for the \Hii bubble to be trapped around the BH, however, the
condition $r_\mr{HII}>r_\mr{B}$ must be satisfied
\citep{Inayoshi:2016ac}.  Otherwise, the gas originally in between
$r_\mr{HII}$ and $r_\mr{B}$ would accumulate around the periphery of the
\Hii bubble. This leads to the enhancement of density $n_\mr{HII}$ and
thus to shrinkage of the bubble. In the end, the \Hii bubble disappears
and radiation feedback no longer affects the accretion. At that time,
the accretion rate returns to the original Bondi value for the neutral
gas $\dot{M}_\mr{B}$, instead of that for the \Hii bubble
$\dot{M}_\mr{B,HII}$.  For this to happen, a system with a massive BH
and/or dense ambient medium, namely $(M_\mr{BH}/10^4
M_\odot)(n_\infty/10^5\,\mr{cm^{-3}}) \gtrsim 1$
\citep{Inayoshi:2016ac}, is required when the BH luminosity is close to
the Eddington value as $L\approx L_\mr{E}$.

Recall that the above argument is based on the assumption of the
spherical symmetry, which should be modified in realistic situations
with anisotropic BH irradiation. We expect that the flow through
shadowed equatorial regions, if exist, enhance the accretion rate.  In
what follows, we will see what kind of the flow structure appears for
such cases by using numerical simulations.

\section{NUMERICAL METHOD}
\label{sec:num-method}

\begin{figure*}
\centering \includegraphics[trim=0 0 0 0, width=17cm,
clip]{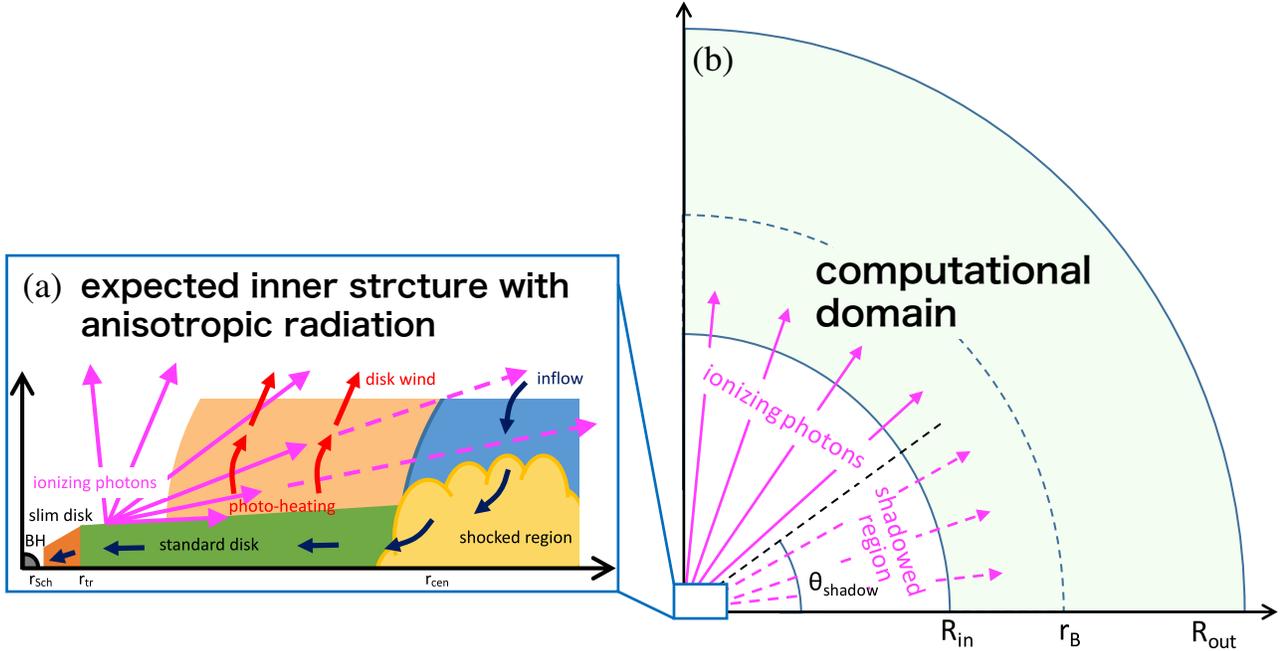} \caption{ Framework of our two-dimensional
radiation hydrodynamics simulations for gas accretion on to BHs: (a) the
envisioned structure of the circum-BH disc, the anisotropic radiation
from which is described by a subgrid model; and (b) the computational
domain with the central sink from which the ionizing photons are
injected following the prescription given by the subgrid model.  In
panel (a), ionizing photons are emitted from an inner hot part of the
disc.  We suppose super-Eddington accretion for this picture, i.e.,
$\dot{M}/\dot{M}_\mr{E}>1$, and a slim disc appears in the innermost
part accordingly.  A standard disc extends to the centrifugal radius
$r_\mr{cen}$.  In the outer part, a wind from the disc surface and/or
swollen shocked layer near the outer disc edge (also see the text) block
the ionizing photons to cast a shadow which has an opening angle
$\theta_\mr{shadow}$.  } \label{fig:acc_whole}
\end{figure*}

We study accretion of primordial gas on to BHs under anisotropic
radiation by performing a series of 2D RHD simulations
(Sec.~\ref{sec:rhd_sim}).  Specifically, we solve the dynamics of the
flow around the Bondi radius (see Fig.~\ref{fig:acc_whole}b), where the
accretion rate on to the BH and disc system is determined.  We mask the
inner circum-BH accretion disc (see Fig.~\ref{fig:acc_whole}a) by the
central sink region and inject ionizing photons at the inner boundary
$R_{\rm in}$ according to a simple parametric sub-grid model that
represents various directional dependences of BH irradiation
(Sec.~\ref{sec:radiation}).

\subsection{Two-dimensional radiation hydrodynamics simulations}
\label{sec:rhd_sim}

We use a modified version of the public multi-dimensional
magneto-hydrodynamics code {\tt Pluto 3.0} \citep{Mignone:2007aa}, which
has been applied to studies on the present-day high-mass star formation
\cite[e.g.,][]{Kuiper:2010aa, Kuiper:2010ab, Kuiper:2011aa,
Kuiper:2013aa} and Pop III star formation \citep{Hosokawa:2016aa}.

Here, we have tuned the code used for the Pop III star formation
\citep{Hosokawa:2016aa} to fit our study of the BH accretion.  As in
\citet{Kuiper:2010aa}, we adopt a 2D polar coordinate system assuming
the axial symmetry.  We calculate only the gravity of the central BH and
neglect the gas self-gravity, as in the previous studies
\citep[e.g.,][but also see Li
2011]{Park:2011aa,Milosavljevic:2009ab,Inayoshi:2016ac}.  We assume that
the outer edge of the accretion disc, i.e., the centrifugal radius
$r_{\rm cen}$, is much smaller than the sink radius $R_\mr{in}$.  We
thus ignore the angular momentum of the flow in the computational
domain.  Other modifications we have added are summarized as follows.

\subsubsection{Chemical and thermal processes}
\label{sec:chemistry}

To solve the chemical and thermal processes, we use the same methods
developed in \cite{Hosokawa:2016aa} with several modifications.  Unlike
in \cite{Hosokawa:2016aa}, we omit $\mr{H}_2$ chemistry assuming that
$\mr{H}_2$ is completely photo-dissociated by the central FUV
irradiation.\footnote{For a case with $M_\mr{BH}=10^3\,M_\odot$ and
$L=L_\mr{E}$ with the spectrum $L_\nu \propto \nu^{-1.5}$, the specific
FUV intensity at $r_\mr{B}$ is $J_{21}\sim 10^9$ (in units of
$10^{-21}\cmr{erg\,s^{-1}\,Hz^{-1}\,sr^{-1}\,cm^{-2}}$), while the
critical intensity for totally suppressing $\mr{H_2}$ formation in
atomic cooling halos is $J_{21,cr}\sim 10^3$ \citep[see,
e.g.,][]{Sugimura:2014aa}.}  We have added the $\mr{He}$ chemistry,
since hard UV photons from BH accretion discs create a large helium
photoionized region embedded in an \Hii region.

In summary, we solve the chemical network with six species: $\mr{H}$,
$\mr{H^+}$, $\mr{e}$, $\mr{He}$, $\mr{He^+}$, and $\mr{He^{2+}}$, which
consists of the following chemical processes: photoionization of
$\mr{H}$, $\mr{He}$ and $\mr{He^+}$; collisional ionization of $\mr{H}$,
$\mr{He}$ and $\mr{He^+}$; recombination of $\mr{H^+}$, $\mr{He^+}$ and
$\mr{He^{2+}}$.  Accordingly we consider the following thermal
processes: photoionization heating of $\mr{H}$, $\mr{He}$, and
$\mr{He^+}$; recombination cooling of $\mr{H^+}$, $\mr{He^+}$, and
$\mr{He^{2+}}$; excitation cooling of $\mr{H}$, $\mr{He}$, and
$\mr{He^+}$; collisional ionization cooling of $\mr{H}$, $\mr{He}$, and
$\mr{He^+}$; free-free cooling of $\mr{H}$, $\mr{He}$, and
$\mr{He^{+}}$; Compton cooling by cosmic microwave background (CMB)
photons.  Complete lists of our adopted chemical and thermal processes
are available in Appendix~\ref{sec:chem_detail}.

We turn off the cooling when the temperature falls below $10^4\cmr{K}$,
as in the previous 2D simulations \citep[e.g.,][]{Park:2011aa}.  We
neglect secondary ionization and heating caused by X-ray photoionization
\citep{Shull:1979ab,Shull:1985aa,Ricotti:2002aa}.  We have confirmed
with test calculations that these processes hardly affect the gas
dynamics though the ionization degree is only slightly enhanced just
outside the \Hii bubble.

\subsubsection{Transfer of ionizing photons}
\label{sec:rad_transfer}

As in \citet{Hosokawa:2016aa}, we only solve the transfer of ionizing
photons directly coming from the central accretion disc.  Diffuse
recombination photons are considered by way of the on-the-spot
approximation.  The radiation transfer is successively solved with the
chemistry from the innermost cell, where photons are injected according
to the sub-grid radiation model (see Sec.~\ref{sec:radiation} below).
We do not consider the absorption between the radiation source and the
inner boundary, which is currently masked by the sink cell.

Regarding the transport of ionizing photons, we have made the following
major updates. First, we solve the frequency-dependent transfer with 128
logarithmically-spaced frequency bins between $13.6\cmr{eV}$ and
$1\cmr{keV}$, to consider the photoionization of $\mr{H}$, $\mr{He}$ and
$\mr{He^+}$ with different threshold energies.  Second, we consider the
radiation pressure via Thomson scattering and photoionization.  As will
be seen in Section~\ref{sec:result}, the radiation pressure becomes
important when the luminosity exceeds the Eddington limit.

We simply assume that photons with energy below $13.6\cmr{eV}$ freely
escape from the system \citep{Park:2011aa}.  Although the radiation
pressure of accumulated Ly$\alpha$ photons would affect the gas dynamics
in spherically symmetric systems, it is probably not the case in
realistic systems with channels for Ly$\alpha$ photons to escape
\citep[e.g.,][]{McKee:2008aa,Milosavljevic:2009aa}.

\subsection{Subgrid model for the irradiation by BH}
\label{sec:radiation}

\begin{figure}
\centering \includegraphics[trim=0 0 0 0, width=7cm,
 clip]{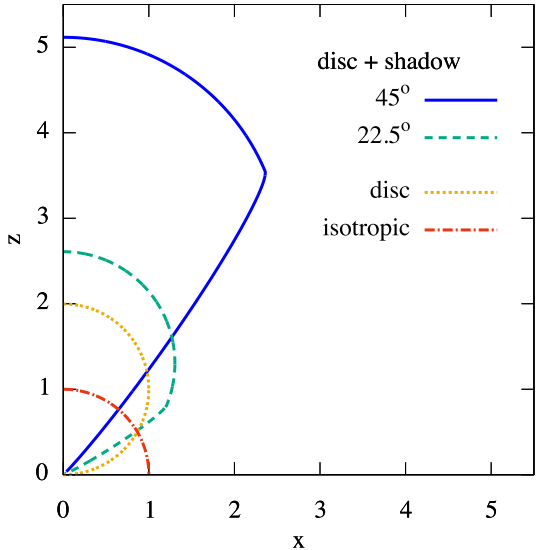} \caption{The anisotropy factor
 $\mathcal{F}(\theta)$ of the BH irradiation given by
 equation~\eqref{eq:5}.  The different curves represent different cases
 considered: isotropic radiation and disc radiation with and without the
 shadowing effect.  For the case with the disc radiation with the
 shadowing effect, we set $\theta_\mr{shadow}=45^\circ$ and
 $22.5^\circ$.  The radial extent $r$ for each angle $\theta$ represents
 the value of $\mathcal{F}(\theta)$ in that direction.  }
 \label{fig:dirdep}
\end{figure}

In our model of anisotropic BH irradiation, we assume that ionizing
photons are emitted from the inner hot part of a circum-BH accretion
disc but a portion of them are absorbed (or scattered) by outer
structures (see Fig.~\ref{fig:acc_whole}a).  We inject ionizing photons
at the inner boundary depending on the inflow rate into the sink cell,
according to the model described here.  We first describe the structure
of the BH accretion disc in Section~\ref{sec:acc_disc}, which motivates
our subgrid model. Then we give the expressions for the luminosity and
directional dependence in Sections~\ref{sec:efficiency} and
\ref{sec:direction_dep}, respectively.

\subsubsection{BH accretion disc with shadowing effect}
\label{sec:acc_disc}

Fig.~\ref{fig:acc_whole}(a) shows the expected inner structure including
a BH accretion disc that motivates our subgrid model.  Below we explain
the inner and outer parts of the structure presented in
Fig.~\ref{fig:acc_whole}(a) in this order.  We also describe resulting
directional dependences of the BH irradiation.

In the inner part, we see that the circum-BH disc consists of the two
different types of accretion discs.  One is the innermost geometrically
thick (aspect ratio $\sim 1$) slim disc appearing inside the photon
trapping radius $r_\mr{tr} \equiv (\dot{M}/\dot{M}_\mr{E})\,r_\mr{Sch}$,
where the cooling via radial advection balances with the viscous heating
\citep[e.g.,][]{Begelman:1978aa,Abramowicz:1988aa}.  The other is the
geometrically thin (aspect ratio $\ll 1$) standard accretion disc
appearing outside $r_\mr{tr}$, where the radiative loss from the disc
surfaces is the main cooling process \citep[e.g.,][]{Shakura:1973aa}.
When $\dot{M}/\dot{M}_\mr{E} < 1$, the slim disc disappears and the
standard accretion disc extends all the way to the inner disk edge.  We
model the luminosity based on this consideration in
Section~\ref{sec:efficiency}.  Since the surface temperature of the disc
increases with decreasing the radius $r$, ionizing photons mostly come
from the hot innermost part.

In the outer part, a disc wind might be launched from the disc surface
photo-heated by the high-energy photons from the inner region
\citep[see][for line-driven disc wind of
AGNs]{Proga:2000aa,Proga:2004aa,Nomura:2013aa,Nomura:2016aa}. In
addition, around the outer edge of the disc, the vertically falling flow
due to the centrifugal barrier might collide with the one coming from
the opposite side of the equatorial plane and form a shocked region.
Fig.~\ref{fig:acc_whole}(a) depicts these structures, both of which can
absorb (or scatter) the ionizing photons coming from the inner part,
forming a shadowed region behind them.  We assume that the shadowing
effect considered here is caused by the outer disc structures, and thus
the appearance of a slim disc is not essential in forming the shadowed
region. The outer structure of the disc is highly uncertain and probably
varies depending on the BH mass, accretion rate, angular momentum,
metallicity of inflowing gas, etc..  In Section~\ref{sec:direction_dep},
we model the shadowing effect with a simple parametric fashion.

\subsubsection{Luminosity}
\label{sec:efficiency}

In our simulation, we determine the luminosity of the BH radiation $L$
depending on $\dot{M}$ evaluated at the inner boundary at each time
step.  To model the luminosity, we adopt the fitting formula
\citep{Watarai:2000aa},
\begin{align}
 L = 
\begin{cases}
\ds 2\, L_\mr{E}\,\left[1+\ln\left(\frac{\dot{m}}{20}\right)\right] & \dot{m} > 20\\
\ds 0.1\, L_\mr{E}\, \dot{m}  & \dot{m} < 20
\end{cases}\,,
\label{eq:3}
\end{align}
where $\dot{m}\equiv \dot{M}/\dot{M}_\mr{E}$.  \cite{Watarai:2000aa}
obtained this formula by fitting the dependence of the luminosity on the
accretion rate in the 1D stationary disc model, taking into account the
(dis-)appearance of the slim disc depending on $\dot{M}$ .  When
$\dot{M}$ is low ($\dot{m} < 20$), the radiative efficiency is fixed at
10\%, which agrees with that of the standard disc.  For rapid accretion
with $\dot{m} > 20$, the second term $2 L_\mr{E}\, \ln(\dot{m}/20)$
represents the luminosity from the innermost slim disc, where the photon
advection reduces the radiative efficiency. Note that the luminosity $L$
increases logarithmically with $\dot{M}$ and can even exceed $L_\mr{E}$
because a large fraction of the emitted photons escape from the disc
surfaces in vertical directions \citep[see
e.g.,][]{Abramowicz:1988aa,Watarai:2000aa,Ohsuga:2005aa,Jiang:2014aa,Sc-adowski:2016aa}.
The first term $2 L_\mr{E}$ corresponds to the luminosity from the outer
standard disc in $r > r_\mr{tr}$ (see Fig.~\ref{fig:acc_whole}a), given
approximately by the energy generation rate due to the gravitational
energy released by $r_\mr{tr}$, $G\dot{M}/r_\mr{tr} \sim L_\mr{E}$
\citep[e.g.,][]{Begelman:1978aa,Kato:1998aa}.

The spectrum of the BH radiation is simply assumed to be the power-law
with $L_\nu\propto \nu^{-1.5}$ for $h\nu>13.6\cmr{eV}$, where
$L=\int_{h\nu>13.6\mr{eV}}L_\nu d\nu$, as often assumed in the
literature
\citep[e.g.,][]{Park:2011aa,Park:2012aa,Milosavljevic:2009ab}.
\cite{Park:2011aa} have shown that the qualitative properties of
accretion do not depend on the spectral shape.

\subsubsection{Directional dependence}
\label{sec:direction_dep}

We inject ionizing photons at the inner boundary with the directional
dependence described below.  Specifically, we multiply the anisotropy
factor $\mathcal{F}(\theta)$ normalized as $\int
\mathcal{F}(\theta)d\Omega=4\pi$ with an isotropic radiation flux
$L/4\pi R_\mr{in}^2$ at the inner boundary $R_\mr{in}$.  With this
definition, $\mathcal{F}(\theta) = 1$ represents the isotropic radiation
(Fig.~\ref{fig:dirdep}).  We use the latitudinal angle $\theta$ defined
as the angle measured from the equatorial plane for our convenience.

Motivated by the expected disc structure described in
Section~\ref{sec:acc_disc} (also see Fig.~\ref{fig:acc_whole}a), we
model $\mathcal{F}(\theta)$ as
\begin{align}
 \mathcal{F}(\theta) &= C\, f_\mr{disc}(\theta)\, f_\mr{shadow}(\theta)\,,
\label{eq:5}
\end{align}
where $C$ is the normalization factor. In this expression, the inner
anisotropy factor $f_\mr{disc}$ that represents the directional
dependences of the radiation emitted from the inner part of the disc is
multiplied by the outer one $f_\mr{shadow}$ to take into account the
outer shadowing effect.

For the inner anisotropy factor $f_\mr{disc}$, we simply assume
\begin{align}
f_\mr{disc}(\theta)\propto\sin\theta\,,
\label{eq:11} 
\end{align}
which corresponds to radiation from an infinitely thin disc (recall that
we define $\theta$ as the angle from the equatorial plane).  Although
numerical simulations suggest somewhat steeper $\theta$-dependence
especially in the polar directions
\citep[e.g.,][]{Ohsuga:2005aa,Sc-adowski:2016aa}, such deviations cause
little effects on our results because the mass accretion predominantly
occurs through the infalling region near the equatorial plane.  For the
disc radiation without the outer shadowing effect (i.e.,
$f_\mr{shadow}=1$), the normalized anisotropy factor is
$\mathcal{F}(\theta) = 2\sin \theta$ (Fig.~\ref{fig:dirdep}).

We model the outer anisotropy factor $f_\mr{shadow}$ as
\begin{align}
 f_\mr{shadow}(\theta)
 =  
\begin{cases}
\ds \exp\left[-\left(\frac{\theta-\tilde{\theta}_\mr{shadow}}{\delta\theta}\right)^2\right]
&0 < \theta < \tilde{\theta}_\mr{shadow}\\[0.4cm]
1  &  \tilde{\theta}_\mr{shadow} < \theta < 90^\circ
\end{cases}
\label{eq:10}
\end{align}
where $\tilde{\theta}_\mr{shadow}=\theta_\mr{shadow}+2\,\delta\theta$,
$\theta_\mr{shadow}$ is the opening angle of the shadow, and
$\delta\theta$ the thickness of the transition region.  Here, we assume
$f_\mr{shadow}$ is symmetric about the equatorial plane.  We adopt the
finite transition region setting $\delta\theta = 6^\circ$ to avoid
artificial ionization structure that appears with
$\delta\theta\rightarrow 0$.  Our conclusions are independent of the
arbitrary choice of a small value for $\delta\theta$.  We show
$\mathcal{F}(\theta)$ for the disc radiation with the outer shadowing
effect with $\theta_\mr{shadow}=45^\circ$ and $22.5^\circ$ in
Fig.~\ref{fig:dirdep}.  With the expression given by
equation~\eqref{eq:10}, the outer anisotropy factor begins to decrease
even for $\theta > \theta_\mr{shadow}$, and takes a value of $\sim 0.01$
at $\theta = \theta_\mr{shadow}$.  Although we fix the shadowing profile
$f_\mr{shadow}(\theta)$ during each simulation run for simplicity, it
probably depends on accretion rates in reality.\footnote{Observations of
Galactic stellar BHs support that disc winds and associated shadowed
regions only exist in the high/soft state and disappear in the low/hard
state \citep[e.g., see][]{Ponti:2012aa}.}  In view of large
uncertainties in the shadowing effect, we perform a number of
simulations varying $\theta_\mr{shadow}$ as a free parameter (see
Sec.~\ref{sec:parameters}).

\subsection{Cases considered}
\label{sec:parameters}
  \begin{table*}
   \centering
   \caption{Summary of model parameters and numerical settings.}
   \label{tab:model}
   \begin{tabular}{lccccccccccc} \hline
run&
$M_\mr{BH}\,[M_\odot]$ & $n_\infty\,[\mr{cm^{-3}}]$ &
$\theta_\mr{shadow}^{a}$& 
$N_r\times N_\theta$ & $R_\mr{in}\,[\mr{AU}]$ &
$R_\mr{out}\,[\mr{AU}]$ & $t_\mr{end}\,[\mr{yr}]$
 \\\hline
Di   & $10^3$ & $10^5$ & {\bf isotropic}$^{b}$ & $512\times144$ & $3\times10^2$ & $6\times10^5$ & $5\times10^5$ \\
Ddn   & $10^3$ & $10^5$ & {\bf disc}$^{c}$ &      $512\times144$ & $3\times10^2$ & $6\times10^5$ & $5\times10^5$ \\
Dds$^{d}$   & $10^3$ & $10^5$ & {\boldmath $45^\circ$}           & $512\times144$ & $2\times10^3$ & $3\times10^6$ & $2\times10^6$\\\\
				                                                                       
s075 & $10^3$ & $10^5$ & {\boldmath $33.75^\circ$}        & $256\times72$ & $2\times10^3$ & $3\times10^6$ & $2\times10^6$\\
s050  & $10^3$ & $10^5$ & {\boldmath $22.5^\circ$}         & $256\times72$ & $2\times10^3$ & $3\times10^6$ & $2\times10^6$\\
s025 & $10^3$ & $10^5$ & {\boldmath $11.25^\circ$}        & $256\times72$ & $2\times10^3$ & $3\times10^6$ & $2\times10^6$\\\\
				                                                                       
M1e2 & {\boldmath $10^2$} & $10^5$ & $45^\circ$           & $256\times72$ & $2\times10^2$ & $1.5\times10^6$ & $2\times10^6$\\
M1e4 & {\boldmath $10^4$} & $10^5$ & $45^\circ$           & $256\times72$ & $2\times10^4$ & $2\times10^7$ & $2\times10^7$\\
M1e5 & {\boldmath $10^5$} & $10^5$ & $45^\circ$           & $256\times72$ & $2\times10^5$ & $1\times10^8$ & $5\times10^7$\\\\
				                                                                       
n1e3 & $10^3$ & {\boldmath $10^3$} & $45^\circ$           & $256\times72$ & $2\times10^3$ & $1\times10^7$ & $5\times10^7$\\
n1e4 & $10^3$ & {\boldmath $10^4$} & $45^\circ$           & $256\times72$ & $2\times10^3$ & $6\times10^6$ & $2\times10^7$\\
n1e6 & $10^3$ & {\boldmath $10^6$} & $45^\circ$           & $256\times72$ & $2\times10^3$ & $2\times10^6$ & $2\times10^6$\\\hline
   \end{tabular}\\
\begin{flushleft}
NOTES.\textemdash $^{a}$Disc radiation with shadowing effect is assumed except for Di and Ddn runs; $^{b}$isotropic radiation; $^{c}$disc radiation
without shadowing effect; $^{d}$Dds run is also called s100, M1e3 and n1e5 runs.
\end{flushleft}
  \end{table*}

We perform a set of simulations to see how the directional dependence of
BH irradiation affects the nature of accretion.  Table~\ref{tab:model}
summarizes model parameters and numerical settings adopted for the cases
examined.  In all the cases, we initially set a static and homogeneous
neutral medium with the number density $n_\infty$ and the temperature
$T_\mr{HI}=10^4\cmr{K}$ around a central BH. The BH mass $M_\mr{BH}$ is
fixed constant during the calculation for simplicity.

In Section~\ref{sec:structure}, we perform three high-resolution
simulations, called ``D-series'' (for ``Directional''), with different
types of the directional dependence of the BH irradiation.  For ``Di
run'' (``i'' for ``isotropic''), we assume the isotropic irradiation,
i.e., $f_\mr{disc} = f_\mr{shadow}=1$ in equation~\eqref{eq:5}.  The
anisotropic disc radiation without the outer shadowing effect, i.e.,
$f_\mr{shadow}=1$, is assumed for ``Ddn run'' (``dn'' for ``disc
no-shadow''), and both $\theta$-dependences of $f_\mr{disc}$ and
$f_\mr{shadow}$ are allowed for ``Dds run'' (``ds'' for ``disc
shadow'').  Below we take $\theta_\mr{shadow} = 45^\circ$ as the
fiducial value for the shadow opening angle.  For the other parameters,
we take $M_\mr{BH}=10^3M_\odot$ and $n_\infty=10^{5}\cmr{cm^{-3}}$.
Note that, for this set of $M_\mr{BH}$ and $n_\infty$, previous studies
with isotropic BH irradiation have shown that the accretion rate is
significantly reduced by radiation feedback
(\citealp{Milosavljevic:2009ab,Park:2012aa,Inayoshi:2016ac}; see also
Sec.~\ref{sec:sph_acc}).

In Sec.~\ref{sec:dependence}, we study how the BH accretion changes with
different shadow size $\theta_\mr{shadow}$, BH mass $M_\mr{BH}$, and
ambient density $n_\infty$.  First, to see the
$\theta_\mr{shadow}$-dependence, we perform three simulations of
``s-series'' (for ``shadow'') with different values of
$\theta_\mr{shadow}$ (Sec.~\ref{sec:sdep}).  Specifically, we take
$\theta_\mr{shadow}=11.25^\circ$, $22.5^\circ$, $33.75^\circ$ and
$45^\circ$.  Second, we study the $M_\mr{BH}$-dependence with the
``M-series'', where we take $M_\mr{BH}=10^2$, $10^3$, $10^4$ and
$10^5\,M_\odot$ (Sec.~\ref{sec:Mdep}).  Finally, the
$n_\infty$-dependence is examined with the ``n-series'', where we take
different values of $n_\infty=10^3$, $10^4$, $10^5$ and
$10^6\cmr{cm^{-3}}$ (Sec.~\ref{sec:ndep}).  In the above simulations, we
take the fiducial values of $M_\mr{BH}=10^3M_\odot$,
$n_\infty=10^{5}\cmr{cm^{-3}}$ and $\theta_\mr{shadow} =45^\circ$ unless
otherwise stated.  We discuss the parameters relevant to the growth of
the remnant BHs of Pop III stars in Section~\ref{sec:conclusion}.

For each case, the inner and outer boundaries $R_\mr{in}$ and
$R_\mr{out}$ are determined in the following way.  We choose small
enough $R_\mr{in}$ to correctly evaluate $\dot{M}$.  To be more
specific, $R_\mr{in}$ is taken to be much smaller than the Bondi radius
for a neutral (ionized) gas when the dominant component of the accreting
gas is neutral (ionized). We choose large enough $R_\mr{out}$ to keep an
\Hii bubble within a simulation region.  We only allow the flow going
out of the computational domain at the inner boundary at $R_\mr{in}$,
where $\dot{M}$ is evaluated.  Across the outer boundary at
$R_\mr{out}$, however, both the inflow and outflow are allowed. In the
angular direction, the computational domain is $0<\theta<90^\circ$ under
the assumption of the equatorial symmetry.

The grid numbers are taken to be $N_r\times N_\theta=512\times144$ and
$256\times 72$ for the high- and medium-resolution simulations,
respectively (see Table~\ref{tab:model}).  In order to simultaneously
resolve the Bondi and Str\"omgren radii, which are different typically
by $3-4$ orders of magnitude, we increase the radial cell size $\Delta
r$ with the fixed size ratio $\Delta r_i/\Delta r_{i-1}$ $(>1)$.  We set
$\Delta r_1 = 0.1 R_\mr{in}$ at the inner boundary.  The grids in the
angular direction are homogeneously distributed over
$0<\theta<90^\circ$, and thus the grid size is $\Delta\theta =
90^\circ/N_\theta$.

We have tested the convergence of the numerical results by varying the
grid numbers or inner boundary radius (see
Appendix~\ref{sec:res_check}).  We follow the evolution over the
duration $t_\mr{end}$, until the accretion reaches a steady state in Di
and Ddn runs, or until $\dot{M}$ reaches almost constant in the other
runs.

\section{Results}
\label{sec:result}

\subsection{Structures of flows}
\label{sec:structure}
\begin{table}
   \centering \caption{Summary of the results in Sec.~\ref{sec:structure}}.
  \label{tab:d-model}
   \begin{tabular}{lccc} \hline
run&
subgrid radiation type& 
$\theta_\mr{inflow}(r_\mr{B})$$^{a}$ &
$\dot{M}/\dot{M}_\mr{B}$$^{d}$
 \\\hline
Di  & isotropic     & ...$^{b}$   & $0.17\%^{e}$\\
Ddn  & disc  & ...$^{c}$   &$0.13\%^{e}$ \\
Dds  & disc $+$ shadow   & $40^\circ$ & $59\%^{f}$\\\hline
   \end{tabular}\\
\begin{flushleft}
NOTES.\textemdash 
$^{a}$opening angle of equatorial neutral inflow region at $r_\mr{B}$ (see text);
$^{b}$no equatorial neutral region;
$^{c}$equatorial neutral region does not reach $R_\mr{in}$;
$^{d}$accretion rate normalized by Bondi one;
$^{e}$averaged between $t=4\times 10^5\cmr{yr}$ and $5\times 10^5\cmr{yr}$;
$^{f}$evaluated at the end of simulation.
\end{flushleft}
  \end{table}

\begin{figure}
\centering \includegraphics[trim = 0 0 0 0,
 clip,width=8cm]{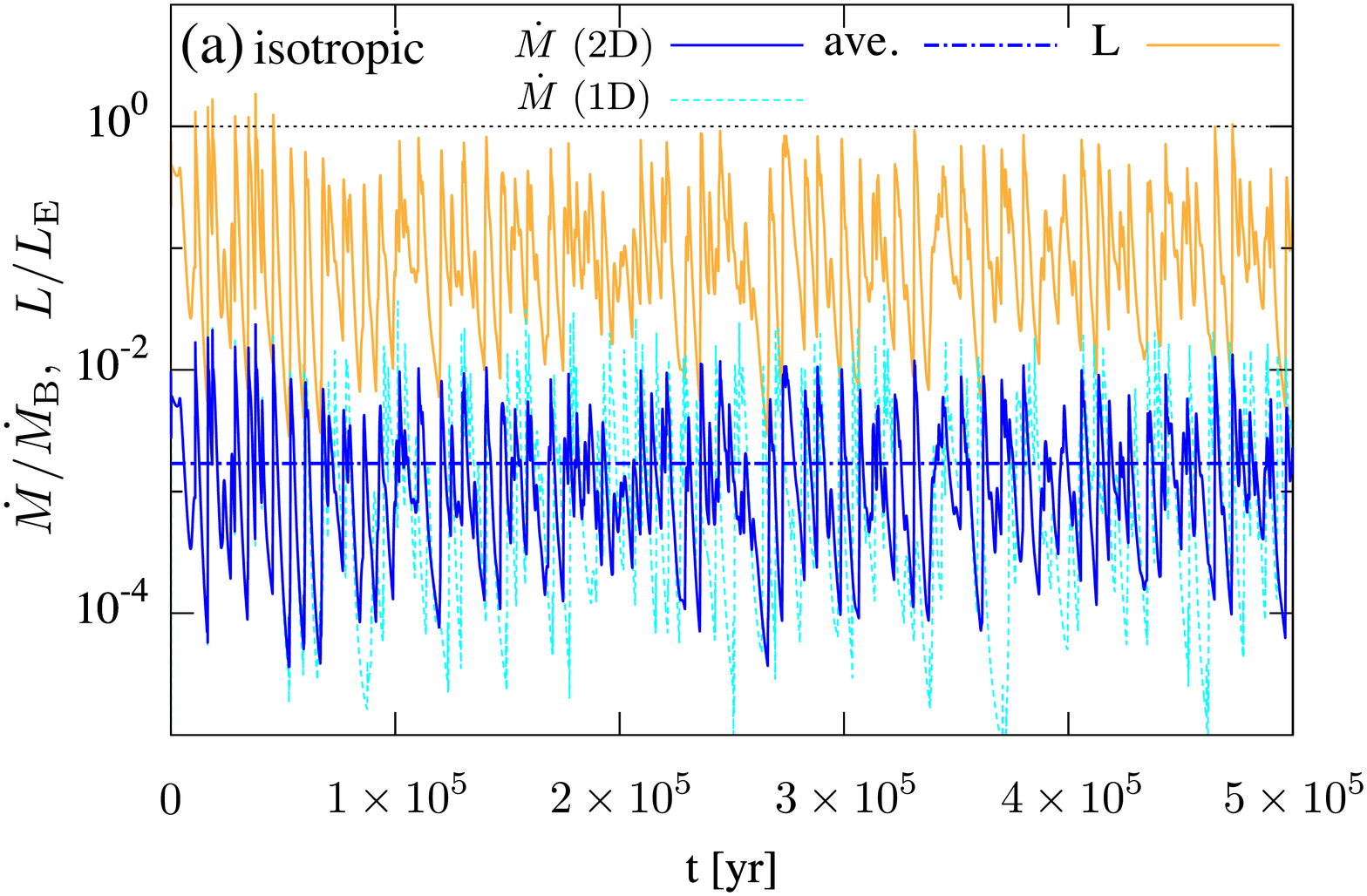}\vspace{0.3cm} \includegraphics[trim
 = 0 0 0 0, clip,width=8cm]{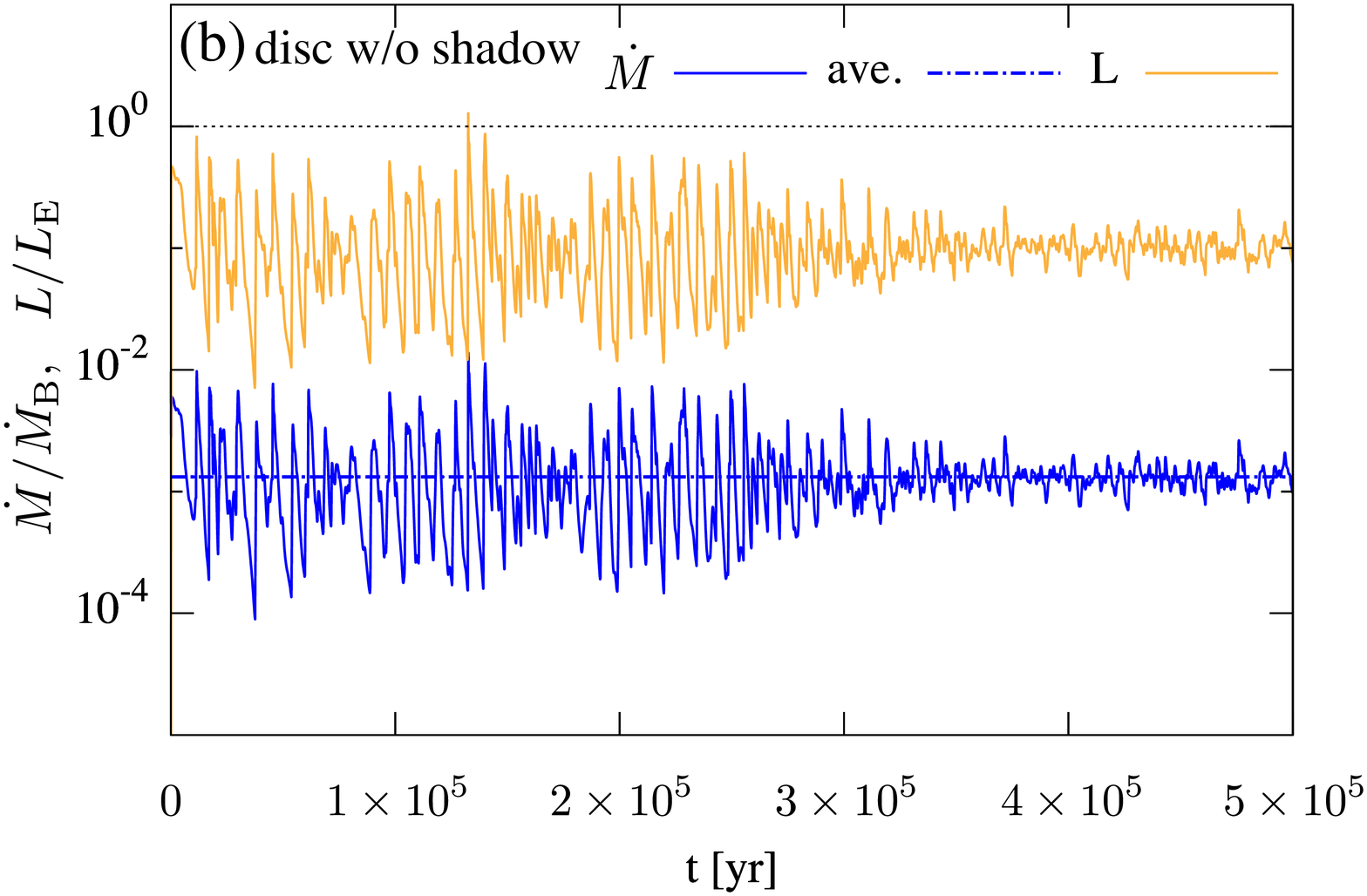}\vspace{0.3cm}
 \includegraphics[trim = 0 0 0 0, clip,width=8cm]{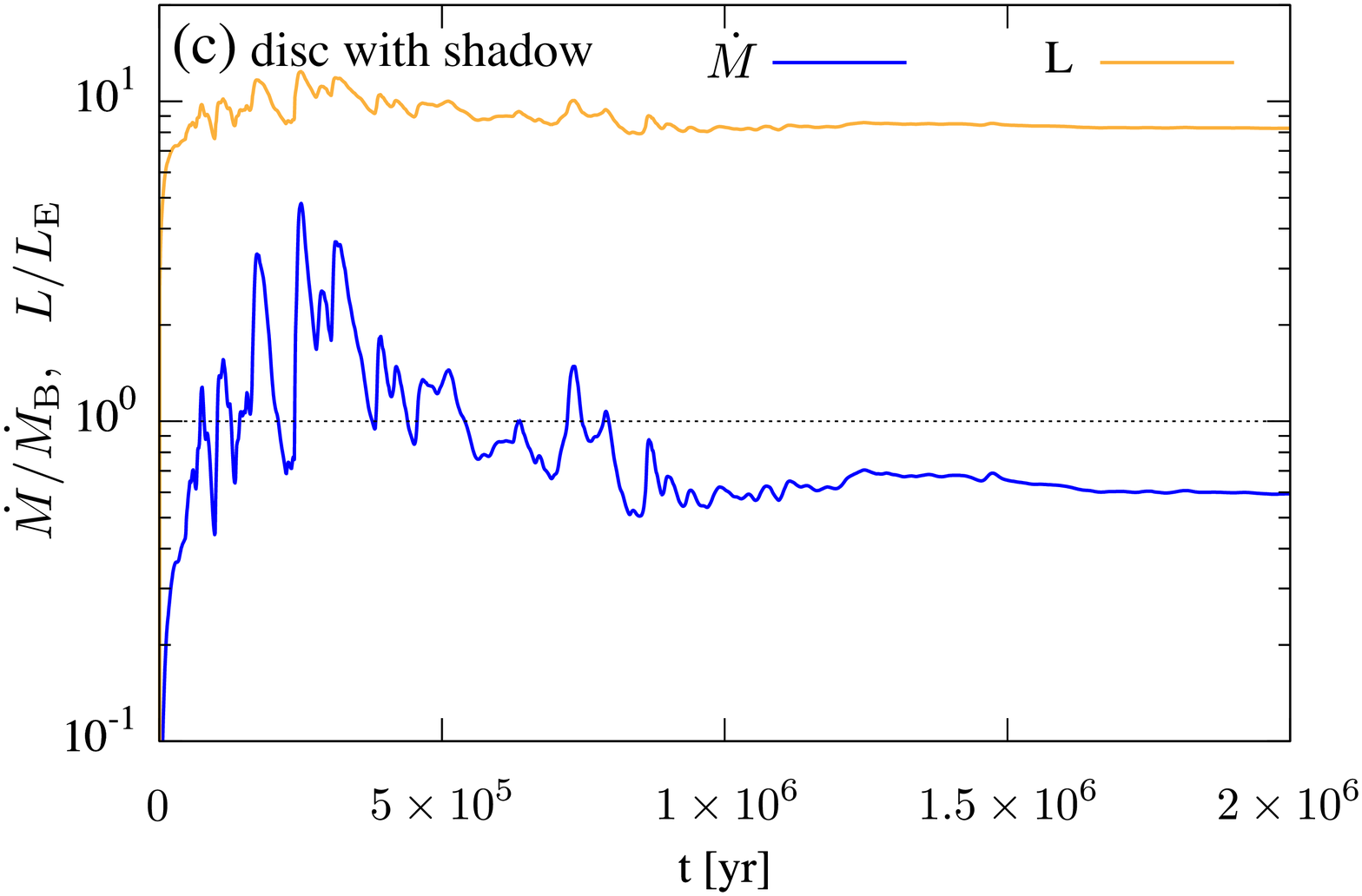}
 \caption{Time evolution of the accretion rate $\dot{M}$ and the
 luminosity $L$ for D-series: (a) Di (isotropic), (b) Ddn (disc) and (c)
 Dds (disc $+$ shadow) runs.  We normalize $\dot{M}$ and $L$ by the
 Bondi rate $\dot{M}_\mr{B}$ and the Eddington luminosity $L_\mr{E}$,
 respectively.  In (a) and (b), $\dot{M}$ (solid blue), $\dot{M}$ on
 average (dot-dashed blue) and $L$ (solid orange) are plotted, while
 $\dot{M}$ in the 1D calculation (dashed cyan) is plotted as well in
 (a).  In (c), $\dot{M}$ (solid blue) and $L$ (solid orange) are
 plotted, with the different vertical and horizontal scales from those
 used in (a) and (b). }  \label{fig:mdot}
\end{figure}

In this section, we perform the simulations of ``D-series'', in
order to see how the flow structure changes with different directional
dependences of the radiation fields.  The basic results for
these cases are summarized in Table~\ref{tab:d-model}.

\subsubsection{Case with isotropic radiation}
\label{sec:spherical_sym}
\begin{figure}
\begin{center}
 \includegraphics[trim = 0 0 0 0, clip, width=7cm]{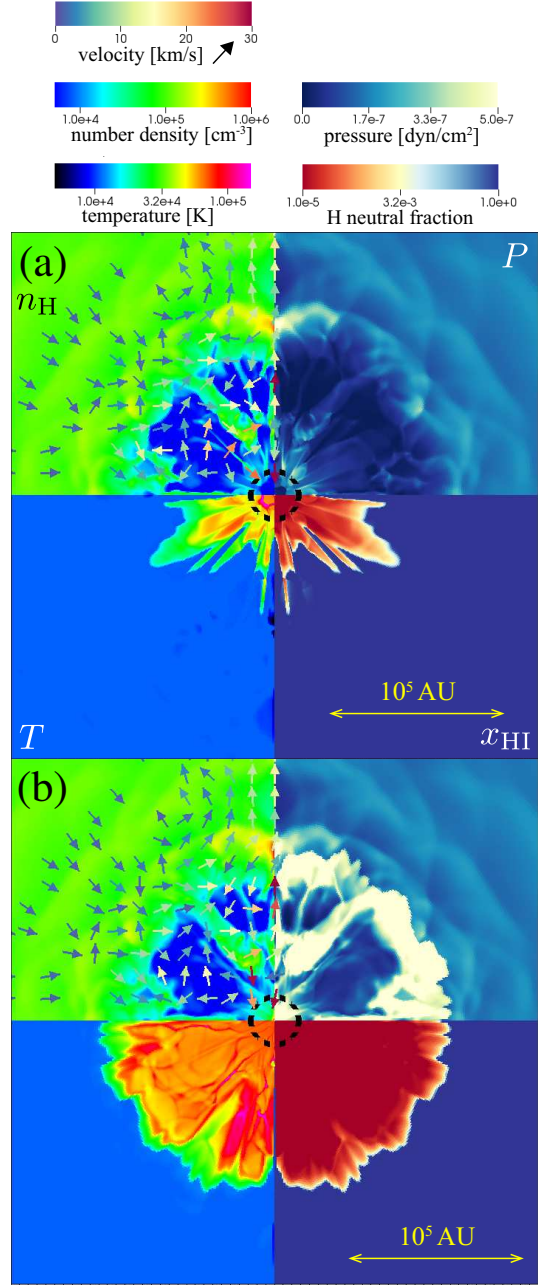}
 \caption{Structures of accretion flow (a) before and (b) after an
 accretion burst for the case with isotropic radiation.  In each
 panel, the four quadrants (clockwise from top left) represent the
 number density $n_\mr{H}\, [\mr{cm^{-3}}]$, the pressure $P\,
 [\mr{dyn\,cm^{-2}}]$, the neutral fraction of hydrogen $x_\mr{H}$ and
 the temperature $T\, [\mr{K}]$, while the arrows represent the velocity
 vector $\mathbf{v}$ only for $|v|>1\cmr{km s^{-1}}$.  The Bondi radius
 $r_\mr{B}$ for the ambient neutral gas is shown as a dashed black circles.}
 \label{fig:snap_Di}
\end{center}
\end{figure}

We first describe ``Di run'', for which we assume the isotropic BH
irradiation with $M_\mr{BH}=10^3M_\odot$ and
$n_\infty=10^5\cmr{cm^{-3}}$.  The behaviour of the accretion flow is
qualitatively the same as those obtained in the earlier works
\citep{Milosavljevic:2009ab,Park:2011aa,Park:2012aa}, and so we refer
the readers to the above literature for full details.

Fig.~\ref{fig:mdot}(a) shows the time evolution of the accretion rate
and luminosity in our 2D simulation, along with the result of 1D
calculation with the same parameter set.  As explained by
\cite{Park:2011aa}, the accretion rate oscillates by repeating the
following three phases: (a) high thermal pressure of the hot \Hii bubble
suppresses the gas inflow and forms a dense shell of the swept up
neutral gas; (b) the accretion rate, and hence the luminosity, decreases
because the density of the \Hii bubble decreases due to the bubble
expansion and/or gas inflow into the sink, leading to the contraction of
the \Hii bubble with the dense shell; (c) an accretion burst caused by
the collapse of the shell dramatically increases the luminosity and
revive the large \Hii bubble again.  The interval time between bursts
roughly corresponds to the sound crossing time across the \Hii bubble.
The accretion history in the 2D simulation is identical to the 1D result
in an early stage ($t\lesssim 5\times10^4\cmr{yr}$), but deviates from
it later on because the spherical symmetry breaks down due to the growth
of numerical perturbations by the instability of expanding ionization
front
\citep[e.g.,][]{Garcia-Segura:1996aa,Whalen:2008ac,Whalen:2008aa,Park:2014ab}.
Although the qualitative features are similar in the 1D and 2D cases,
the accretion variability is slightly weaker in the latter case.  The
peaks of accretion burst in different directions are smoothed out
because they are not exactly synchronized in the 2D case.

The average accretion rate between $t=4\times 10^5\cmr{yr}$ and $5\times
10^5\cmr{yr}$ is only $\dot{M}=1.7\times10^{-3}\dot{M}_\mr{B}$.  Such a
low rate is consistent with the Bondi rate in the ionized medium,
$\dot{M}_\mr{B,HII} \sim 1 \times 10^{-3} \dot{M}_\mr{B}$ (see
Sec.~\ref{sec:sph_acc}).  Our result is in good agreement with the
previous ones by \citet{Milosavljevic:2009ab} and \citet{Park:2012aa},
who provided $\dot{M}\sim 2\times10^{-3}\dot{M}_\mr{B}$ and $\dot{M}\sim
10^{-2}\dot{M}_\mr{B}$, respectively.  The differences of a factor of a
few might come from differences in the adopted chemistry, because the
accretion rate is sensitive to the thermal structure within the \Hii
bubble \citep{Park:2011aa,Park:2012aa}.

Fig.~\ref{fig:snap_Di}(a) and (b) show the structures of accretion flows
before and after an accretion burst, respectively.  As explained above,
the \Hii bubble shrinks before the burst (Fig.~\ref{fig:snap_Di}a) and
expands again due to the enhanced luminosity after the burst
(Fig.~\ref{fig:snap_Di}b).  Whereas the Bondi radius $r_\mr{B}$ for the
ambient neutral gas is illustrated in the figure, $r_\mr{B,HII}$ for the
ionized medium is, although resolved in our simulations, too small to be
shown. The velocity field of the ionized gas does not exhibit a
systematic inflow but subsonic turbulent structure since the gas
pressure dominates the gravity outside the Bondi radius.  These
snapshots are also very similar to those shown in the previous works
\citep{Milosavljevic:2009ab,Park:2011aa,Park:2012aa}.

Note that there appears a thin finger-like structure of the neutral gas
along the $z$ axis in Fig.~\ref{fig:snap_Di}(b).  Since both initial
condition and BH irradiation are spherically symmetric, the flow
patterns should also be spherically symmetric at least in a statistical
sense. Thus, this is an artifact of our 2D simulation, presenting its
limitation.  Any flows toward the $z$ axis inevitably collide each other
on the axis due to the assumed axisymmetry, creating a high density
neutral gas column that shadows the cells behind it.  We expect this
artifact vanishes in future 3D simulations.

\subsubsection{Case with disc radiation without shadowing effect}
\label{sec:disc_rad}

Next, we consider ``Ddn run'', for which we assume the disc radiation
without shadowing effect.  Specifically, we adopt the anisotropy factor
$\mathcal{F}=2\sin\theta$ by taking $f_\mr{shadow}=1$ in
equation~\eqref{eq:5}.

Fig.~\ref{fig:mdot}(b) presents that, as in the case with isotropic
radiation, the accretion rate and luminosity initially show strong
oscillatory behaviours.  However, the oscillation settles down in
$3\times10^5\cmr{Myr}$, after which only weak variability remains.  We
consider that the initial strong oscillation occurs due to the
artificial initial condition of a static homogeneous medium.  The
accretion variability gradually ceases, after which the flow structure
reaches a quasi-steady state.  In this case, burst accretions coming
from different directions cannot be synchronized due to the aspherical
shape of the \Hii bubble created by the anisotropic BH irradiation,
resulting in the less variable accretion rate.  The mean accretion rate
between $t=4\times 10^5\cmr{yr}$ and $5\times 10^5\cmr{yr}$, when the
oscillation has already abated, is
$\dot{M}=1.3\times10^{-3}\dot{M}_\mr{B}$.  This low rate is nearly the
same as that in the isotropic irradiation case, and also well
approximated by the Bondi rate from the ionized medium (see
Sec.~\ref{sec:sph_acc}). The inflows through the equatorial neutral
region have little contribution to the accretion rate, as will be seen
below.

\begin{figure}
\centering \includegraphics[width=7cm]{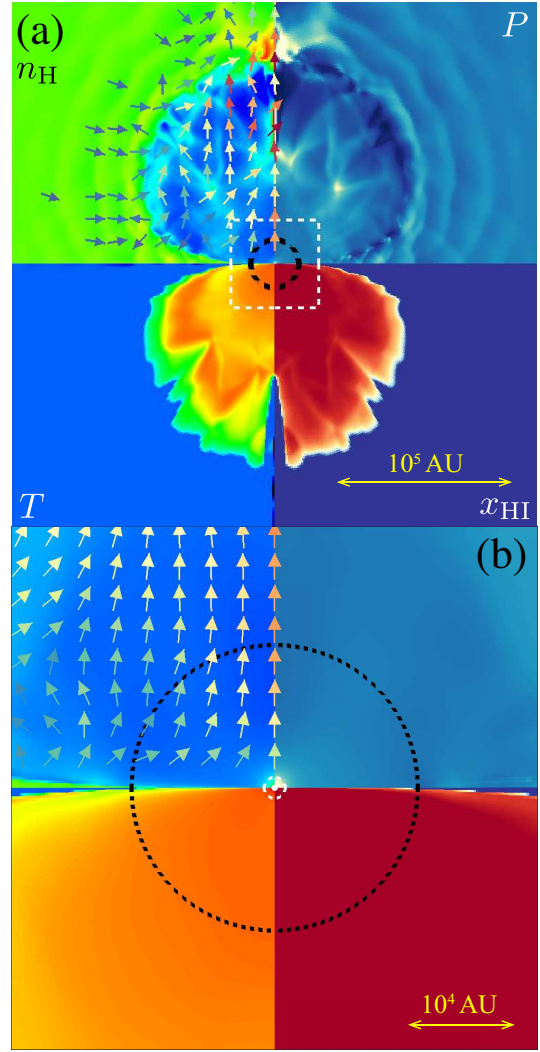}
 \caption{Same as Fig.~\ref{fig:snap_Di} but for the case with disc
 radiation without shadowing effect.  The structure of accretion flow
 at the end of the simulation is shown on the larger and smaller scales
 in panels (a) and (b), respectively.  The white dashed square in panel
 (a) represents the region plotted in (b).  In addition to the Bondi radius for
 the ambient neutral gas ($r_\mr{B}$, dashed black) shown in both
 panels (a) and (b), that for the ionized medium ($r_\mr{B,HII}$, dashed
 white) is shown in (b).}  \label{fig:snap_Dd}
\end{figure}

Fig.~\ref{fig:snap_Dd} shows the structure of the accretion flow at the
end of the simulation.  The whole \Hii bubble is shown in the upper
panel, while the central region on the scale of $r_\mr{B}$ is enlarged
in the lower panel.  In Fig.~\ref{fig:snap_Dd}(a), we see that the
pressure equilibrium is approximately realized throughout the simulation
region.  The \Hii bubble is squeezed in the equatorial directions
because of the anisotropic irradiation.  The gas within the \Hii bubble
moves upward until colliding with the ambient neutral medium.  A
high-density region near the $z$ axis just outside the \Hii bubble is
again likely to be an artifact, as seen in Sec.~\ref{sec:spherical_sym}.
Hereafter, we will ignore this kind of features since it hardly affects
our conclusion.  In Fig.~\ref{fig:snap_Dd}(b), the gas is ionized in
most of the region except the equatorial thin neutral layer, which
extends inward across the Bondi radius but does not reach the sink.
Most of the outflow within the \Hii region is launched from this
equatorial neutral layer.  In this figure, we do not clearly see the
inflow, which is actually limited to a very central part, because the
Bondi radius for the ionized medium $r_\mr{B,HII}$ is currently much
smaller than the size of the plotted area.

With the current spatial resolution, the thickness of the equatorial
neutral layer is limited by the angular cell size of $\Delta \theta =
0.6^\circ$.  The ionizing photon flux injected into the cells closest to
the equatorial plane ($0 \leq \theta \leq \Delta\theta$) is reduced to
$\sim 0.01$ of the angle-averaged value for the assumed anisotropy.
This flux is, however, still large enough to make the \Hii region extend
beyond the sink radius.  In fact, the Str\"omgren radius with $n=
n_\infty =10^5\cmr{cm^{-3}}$ and $L = 10^{-3}L_\mr{E}$ gives $r_\mr{HII}
\simeq 7\times10^3\cmr{AU}$ (equation~\ref{eq:13}), which is much larger
than the sink radius $R_\mr{in} = 3 \times 10^2\cmr{AU}$.  The
horizontal extension of the bubble is even larger than this because the
actual bubble density is smaller than $n_\infty$.  If we could perform a
simulation with much higher resolution, a thinner equatorial neutral
region would reach the sink as the less ionizing photon flux is injected
for the smaller $\theta$.  Such a very thin neutral region, however, is
not expected to affect the overall accretion, because the mass that can
be carried through such a very thin region is severely limited and is
further reduced by the mass loss into the \Hii bubble, as will be shown
later (Sec.~\ref{sec:shadow_rad} - \ref{sec:mass_loss_inner}).
Moreover, diffuse recombination photons processed within the bubble,
which are not considered in the current simulations, would eliminate
such a very thin neutral region \citep[see, e.g.,][]{Hollenbach:1994aa,
Tanaka:2013aa}.

\subsubsection{Case with disc radiation with shadowing effect}
\label{sec:shadow_rad}
\begin{figure}
\centering \includegraphics[trim = 0 0 0 0, clip,
 width=7cm]{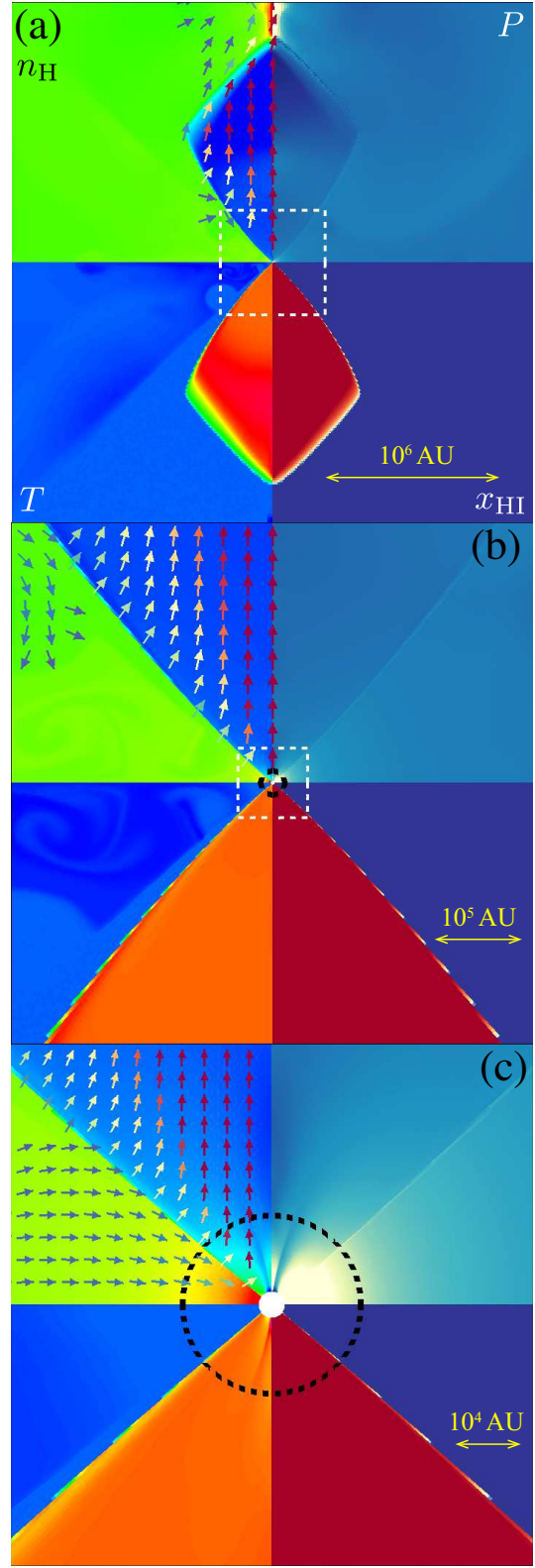} \caption{ Same as
 Fig.~\ref{fig:snap_Di} but for the case with disc radiation with
 shadowing effect.  The structure of the accretion flow at the end of
 the simulation is presented covering the different spatial scales of
 (a) $\sim 10^6$~AU, (b) $\sim 10^5$~AU, and (c) $\sim 10^4$~AU.  The
 white dashed squares in panels (a) and (b) correspond to the plotted
 regions of (b) and (c), respectively.}
 \label{fig:snap_Ds}
\end{figure}

Finally, we describe ``Dds run'', in which the inner disc radiation is
modified by the outer shadowing effect, as
$\mathcal{F}=C\,f_\mr{disc}\,f_\mr{shadow}a$ (equation~\ref{eq:5}).  We
adopt $\theta_\mr{shadow}=45^\circ$ for the outer anisotropy factor
$f_\mr{shadow}$ given by equation~\eqref{eq:10}.

Fig.~\ref{fig:mdot}(c) shows that the accretion rate converges towards
an constant value $\dot{M} \simeq 0.59~\dot{M}_\mr{B}$ in $1\cmr{Myr}$,
which is much higher than the value $\dot{M} \lesssim
2\times10^{-3}~\dot{M}_\mr{B}$ obtained in the former two cases with
isotropic radiation and disc radiation without shadowing effect.  This
accretion rate is also ``super-critical'' and 400 times larger than the
Eddington-limited rate $\dot{M}_\mr{E}$.  By the end of the simulation,
the luminosity also converges to $L \simeq 8L_\mr{E}$, i.e., a
super-Eddington luminosity realized by the high accretion rate.

Fig.~\ref{fig:snap_Ds} shows the structure of the accretion flow at the
end of the simulation.  The whole \Hii bubble is shown within the large
plotted area of Fig.~\ref{fig:snap_Ds}(a), while the central regions
over $\sim 10^6$~AU and $\sim 10^5$~AU scales are enlarged in
Figs.~\ref{fig:snap_Ds}(b) and \ref{fig:snap_Ds}(c).  Owing to the
shadowing effect, we see in Fig.~\ref{fig:snap_Ds}(a) that the large
horizontal neutral region cuts into the central part with the \Hii
bubbles bound to the bipolar regions.  Similarly to the case with disc
radiation without shadowing effect (Sec.~\ref{sec:disc_rad}), the gas
within the bipolar \Hii bubbles flows outward and collides with the
ambient neutral medium. The pressure equilibrium is approximately
achieved throughout the simulation region, although the thermal pressure
slightly decreases before the collision because the total pressure
including the ram pressure is balanced.  The size of the bipolar \Hii
bubbles is much larger than in the former cases owing to the much higher
luminosity $L \simeq 8L_\mr{E}$.  Note that the injected radiation is
super-Eddington only in the polar region with $\theta \gtrsim 50^\circ$.
 
Fig.~\ref{fig:snap_Ds}(b) shows that the gas inside the equatorial
neutral region is almost at rest in the pressure equilibrium, while that
on the surfaces is photoevaporated to join the ionized outflow.  In
Fig.~\ref{fig:snap_Ds}(c), where the structure over the scale of the
Bondi radius is presented, the gas flows into the central sink through
the equatorial neutral region.  The density and pressure increase with
decreasing $r$ for $r \lesssim r_\mr{B}$, as expected for the Bondi
flow.  As seen in Fig.~\ref{fig:snap_Ds}(b), the neutral gas is
photoevaporated into the \Hii regions, where the acceleration by the
radiation pressure is stronger than the gravitational pull owing to the
super-Eddington fluxes in the polar directions.  The accretion proceeds
only through the solid angle covered by the equatorial neutral region.
Note that, in our simulations, we neglect possible photoevaporation
outflow coming out from the sink. We discuss it later in
Sec.~\ref{sec:mass_loss_inner}.  In the next section, we investigate the
structure of the flow in more detail.

\subsubsection{Analysis of  flow structure in case with shadowing effect}
\label{sec:anl_modeling}

\begin{figure}
\centering \includegraphics[trim=0 0 0 0, width=7.5cm,
clip]{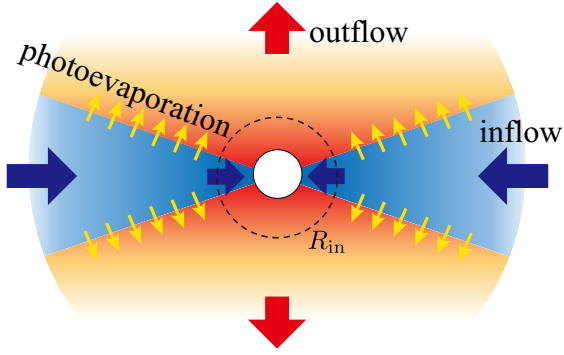} \caption{A schematic view of our analytical
modelling of the flow structure under BH irradiation with the shadowing
effect.  Gas inflows in a Bondi-like fashion in the equatorial neutral
region (blue), while it outflows in the bipolar \Hii regions (red) owing
to the thermal and radiation pressure.  Some of the inflowing gas is
photoionized and supplied into the \Hii regions through the boundaries
between the neutral and \Hii regions, where these two phases of gas are
approximately in pressure equilibrium with each other.  }
\label{fig:anl_model}
\end{figure}

In this section, we develop an analytical model and compare it with our
result to examine the inflow-outflow structure presented in
Fig.~\ref{fig:snap_Ds}. The overall structure of our model is
schematically depicted in Fig.~\ref{fig:anl_model} and can be summarized
as follows: in the equatorial neutral region where ionizing photons
cannot penetrate, the gas inflows in a Bondi accretion fashion; in the
bipolar \Hii regions where ionizing photons heat up the gas via
photoionization, the outflows are launched due to the thermal and
radiation pressure; through their boundaries, the photoevaporating gas
is lost from the neutral region and supplied into the \Hii regions.

We begin with considering the density profiles of the inflow and
outflow.  The radial density profile in the equatorial neutral region
$n_\mr{inflow}(r)$ is well approximated by that of the Bondi solution,
which we further simplify as
 \begin{align}
  n_\mr{inflow}(r) &=
\begin{cases}
\ds n_\infty\left(\frac{r}{r_\mr{B}}\right)^{-3/2}&r < r_\mr{B}\\[0.4cm]
n_\infty&r > r_\mr{B}
\end{cases}\label{eq:9} \,.
 \end{align}
This expression slightly over- and underestimates the obtained density
profile at $r\ll r_\mr{B}$ and $r\sim r_\mr{B}$.  As for the bipolar
ionized outflows, the density profile $n_\mr{outflow}(r)$ can be
estimated by assuming the pressure equilibrium at the conical boundaries
between the neutral and ionized gas,
\begin{align}
n_\mr{outflow}(r)&=n_\mr{inflow}(r)\left(\frac{T_\mr{HI}}{2T_\mr{HII}}\right)\,,
\label{eq:7}
 \end{align}
where $T_\mr{HI} \simeq 10^4\cmr{K}$ and $T_\mr{HII} \simeq
7\times10^4\cmr{K}$ are the temperature in the neutral and ionized
regions, respectively.  Although not very precise, this simple
expression captures the qualitative features of the bipolar ionized
outflows.

\begin{figure}
\centering \includegraphics[width=8.5cm, clip]{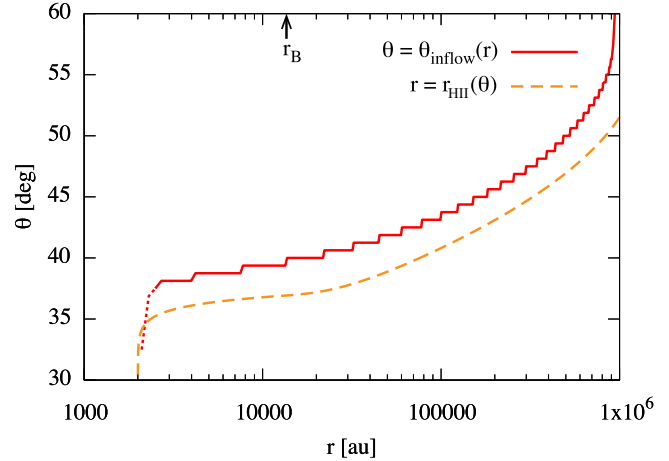}
\caption{The opening angle of the equatorial neutral inflow region
$\theta_\mr{inflow}$ as a function of $r$ (solid red), along with the
angle-dependence of the radius of the {\Hii} region
$r=r_\mr{HII}(\theta)$ (dashed orange; equation~\ref{eq:6}).  The dotted
part of the red line marks a few innermost cells that are artificially
ionized as we neglect the absorption within the sink (also see the
text).  } \label{fig:th_in_Ds}
\end{figure}

Fig.~\ref{fig:th_in_Ds} shows the opening angle of the equatorial
neutral region $\theta_\mr{inflow}$ as a function of $r$. In practice,
we define the neutral region as the region where the ionization degree
of hydrogen is less than $50\%$.  Although $\theta_\mr{inflow}$
decreases as $r$ decreases, the neutral inflow region reaches the inner
boundary at $r = R_\mr{in}$ with a finite angle, unlike in the case with
disc radiation without shadowing effect.  The sharp drop of
$\theta_\mr{inflow}$ around $R_\mr{in} = 2000$~AU is caused by
photoionization of a few innermost cells due to our ignorance of the
consumption of ionizing photons within the sink, although this does not
affect our conclusion.  The opening angle at the Bondi radius,
$\theta_\mr{inflow}(r_\mr{B})=40^\circ$, is similar to the assumed
shadow opening angle $\theta_\mr{shadow} =45^\circ$
(equation~\ref{eq:10}).

In order to estimate $\theta_\mr{inflow}$, we calculate the radius of
the {\Hii} region $r_\mr{HII}$ in each direction $\theta$ with the
modeled density profiles.  The supply rate of ionizing photons per unit
solid angle is given by $\dot{N}_\mr{ion}\,\mathcal{F}(\theta)/4\pi$,
where $\dot{N}_\mr{ion}$ is the total ionizing photon emissivity of the
central accretion disc, given by $\dot{N}_\mr{ion}=L/3h\nu_\mr{T}$ for
the assumed spectral shape of $L_\nu\propto \nu^{-1.5}$. Equating this
supply rate with the recombination rate, we have
\begin{equation}
\frac{L}{3h \nu_\mr{T}}\frac{\mathcal{F}(\theta)}{4 \pi}
= \int_{R_\mr{in}}^{r_\mr{HII}}(\theta)
\alpha_\mr{B}\,n_\mr{outflow}^2\,r^2\,\mr{d}r ,
\label{eq:uve}
\end{equation}
where $n_\mr{outflow}$ is given by equation~\eqref{eq:7}.  Performing
the integration in equation~\eqref{eq:uve}, we finally get
\begin{align}
 r_\mr{HII}(\theta) 
=
\begin{cases}
\ds
R_\mr{in}  \exp\left[
\frac{L\,\mathcal{F}(\theta)\,T_\mr{HII}^2}{3\pi h\nu_\mr{T}\, \alpha_\mr{B}\, n_\infty^2\, r_\mr{B}^3\, T_\mr{HI}^2}
\right]&r_\mr{HII} < r_\mr{B}\\[0.4cm]
\ds
\left[
\frac{L\,\mathcal{F}(\theta)\,T_\mr{HII}^2}
{\pi h\nu_\mr{T}\,\alpha_\mr{B}\,n_\infty^2\,T_\mr{HI}^2} 
- 3r_\mr{B}^3 \ln\left(\frac{r_\mr{B}}{R_\mr{in}}\right)
+ r_\mr{B}^3
\right]^{1/3} \hspace{-2cm}\\[0.4cm]
&r_\mr{HII} > r_\mr{B}
 \end{cases}\,.
\label{eq:6}
\end{align}
Here, we show the relation $r=r_\mr{HII}(\theta)$, or equivalently
$\theta=r_\mr{HII}^{-1}(r)$, in Fig.~\ref{fig:th_in_Ds} with $L =
8.2\,L_\mr{E}$ (Fig.~\ref{fig:mdot}).  We see that equation~\eqref{eq:6}
qualitatively reproduces $\theta_\mr{inflow} (r)$ obtained in the
simulation with a small deviation of a few degrees at each $r$.  Such a
deviation mainly comes from approximate modelling of $n_\mr{outflow}$ in
equation~\eqref{eq:7}.  For example, in an outer part of the \Hii bubble
where helium is not doubly ionized, the gas is no longer heated up to
$7\times10^4\cmr{K}$ by the $\mr{He^+}$ photoionization, resulting in
the higher density than that estimated by equation~\eqref{eq:7} with
$T_\mr{HII} = 7 \times 10^4\cmr{K}$.  Nonetheless, our simple modelling
with equation~\eqref{eq:6} describes the numerical results well.

\begin{figure}
\centering \includegraphics[width=8.5cm, clip]{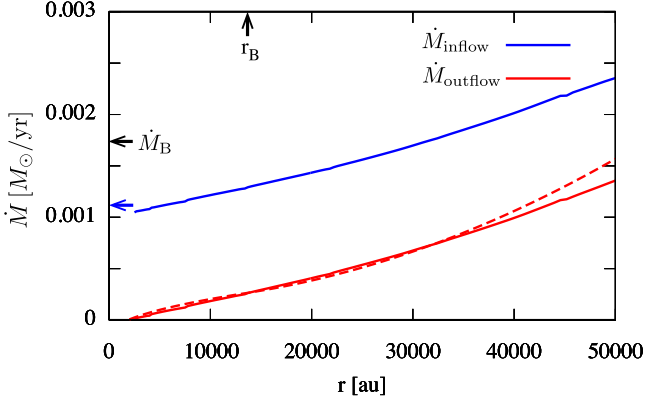}
\caption{The radial dependence of the equatorial inflow rate
$\dot{M}_\mr{inflow}(r)$ (solid blue) and bipolar outflow rate
$\dot{M}_\mr{outflow}(r)$ (solid red), defined in
equation~\eqref{eq:16}.  The blue arrow marks the estimated value of the
inflow rate at the inner boundary $\dot{M}_\mr{inflow}(R_\mr{in})$
(equation~\ref{eq:17}).  The red dashed line represents the analytical
estimate of the outflow rate $\dot{M}_\mr{outflow}(r)$
(equation~\ref{eq:14}).  The values of the Bondi radius $r_\mr{B}$ and
accretion rate $\dot{M}_\mr{B}$ are marked by the black arrows.  }
\label{fig:mdoio_Ds}
\end{figure}

Finally, we investigate the flow rates in the neutral and ionized
regions.  Fig.~\ref{fig:mdoio_Ds} shows the equatorial inflow rate
$\dot{M}_\mr{inflow}$ and bipolar outflow rate $\dot{M}_\mr{outflow}$
through a spherical surface with radius $r$,
\begin{align}
 \dot{M}_\mr{inflow}(r) &=  - 4\pi r^2\int_0^{\theta_\mr{inflow}}  \rho\,v_r\,\cos \theta d\theta\,,\nnmb
\dot{M}_\mr{outflow}(r) &=  4\pi r^2\int_{\theta_\mr{inflow}}^{\pi/2} \rho\,v_r\, \cos \theta d\theta\,,
\label{eq:16}
\end{align}
where $v_r$ is the outward velocity. We have multiplied a factor of two
to take into account the equatorial symmetry.  The net accretion rate
$\dot{M} \equiv \dot{M}_\mr{inflow}-\dot{M}_\mr{outflow}$ is almost
constant with $r$, consistent with a quasi-steady flow structure. The
value of $\dot{M}$ is equal to the inflow rate $\dot{M}_\mr{inflow}$ at
the inner boundary $R_\mr{in}$, where $\dot{M}_\mr{outflow} = 0$ is
imposed as the boundary condition.

We now estimate $\dot{M}_\mr{inflow}(R_\mr{in})$ ($=\dot{M}$) from the
Bondi-like accretion through a solid angle corresponding to the opening
angle $\theta_\mr{inflow}(r_\mr{B})$.  With the solid angle
$\Delta\Omega_\mr{inflow}(r_\mr{B})=4\pi
\,\sin\theta_\mr{inflow}(r_\mr{B})$, we obtain
\begin{align}
 \dot{M}_\mr{inflow}(R_\mr{in})=
\frac{\Delta\Omega_\mr{inflow}(r_\mr{B})}{4\pi}\dot{M}_\mr{B}\,.
\label{eq:17}
\end{align} 
Fig.~\ref{fig:mdoio_Ds} demonstrates that equation~\eqref{eq:17} well
reproduces $\dot{M}_\mr{inflow}(R_\mr{in})$ measured in the
simulation,\footnote{With equation~\eqref{eq:17}, we are now able to
explain why the accretion rate temporarily exceeds the Bondi rate in the
early stage of the simulation in Fig.~\ref{fig:mdot}(c).  In the
beginning, the hot bipolar \Hii bubbles compresses the equatorial
neutral layer. Consequently, the corresponding Bondi-like accretion rate
increases and becomes larger than the original Bondi rate
$\dot{M}_\mr{B}$ even after multiplying the fraction of the neutral
solid angle $\Delta\Omega_\mr{inflow}(r_\mr{B})/(4\pi)$.}  although
slightly overestimated owing to the photoevaporation mass loss.  Note
that the inflow rate $\dot{M}_\mr{inflow}$ increases with $r$ and
exceeds the Bondi rate $\dot{M}_\mr{B}$ at $\sim 3000$~AU.  This is
because the circulation flows are generated to compensate the
photoevaporation mass loss from the equatorial neutral region (see
below) and contribute to the accretion rate together with the Bondi-like
inflow.

The outflow rate can be modeled as the photoevaporation mass loss.  The
mass-loss flux from the surfaces of neutral region is estimated as
$f_\mr{outflow}\,\rho_\mr{b,HII}\, c_\mr{s,HII}$, where $f_\mr{outflow}$
is an $O(1)$ correction factor, $\rho_\mr{b,HII}$ the density at the
bottom of the ionized layer and $c_\mr{s,HII}$ the sound velocity for
ionized gas \citep[e.g.,][]{Hollenbach:1994aa,Tanaka:2013aa}.  The
outflow rate $\dot{M}_\mr{outflow}$ through a given radius $r$ is
obtained by integrating the mass-loss fluxes between $R_\mr{in}$ and $r$
because the outflow is in a quasi-steady state. In reality, additional
mass loss may happen even inside $R_\mr{in}$, as we will discuss in
Sec.~\ref{sec:mass_loss_inner}.  With $\rho_\mr{b,HII} \simeq m_\mr{p}
n_\mr{outflow}$ (equation~\ref{eq:7}) and
$c_\mr{s,HII}=(2T_\mr{HII}/T_\mr{HI})^{1/2}c_\mr{s,HI}$, we obtain
\begin{align}
 \dot{M}_\mr{outflow}(r)
\simeq 4\pi\int_{R_\mr{in}}^r &f_\mr{outflow}\,m_\mr{p}\,c_\mr{s,HII}\,n_\mr{outflow} r' \mr{d}r'\nnmb
=
2\pi\,f_\mr{outflow}\,& m_\mr{p}\, n_\infty\,c_\mr{s,HI}
\left(\frac{2T_\mr{HII}}{T_\mr{HI}}\right)^{1/2}\nnmb
\times&
\begin{cases}
4 r_\mr{B}^{3/2}  \left(r^{1/2}-R_\mr{in}^{1/2}\right)
& r < r_\mr{B} \\
r^2 +  3r_\mr{B}^2 -4r_\mr{B}^{3/2}R_\mr{in}^{1/2}
&
r > r_\mr{B}
\end{cases}\,,
\label{eq:14}
\end{align}
where a factor of two is multiplied in the first equality to take into
account both top and bottom surfaces.  We see in Fig.~\ref{fig:mdoio_Ds}
that modeled $\dot{M}_\mr{outflow}$ with the best-fit value of
$f_\mr{outflow}=0.7$ reproduces the simulation result with remarkable
agreement.

\subsubsection{Possible mass loss from neutral inflow inside the sink}
\label{sec:mass_loss_inner}

As mentioned above, we neglect the possible mass loss from the innermost
part of the flow masked by the sink.  Since the size of the accretion
disc is supposed to be much smaller than the sink radius $R_\mr{in}$
(see Fig.~\ref{fig:acc_whole}), we neglect the centrifugal effect and
assume the similar flow structure extends inward.  As an upper limit for
the mass-loss rate, we evaluate the integral with the same integrand as
equation~\eqref{eq:14} but for the different range of $0< r <
R_\mr{in}$, and obtain
\begin{align}
 \dot{M}_\mr{loss}(<R_\mr{in})
\lesssim 0.1 \left[\frac{R_\mr{in}}{(r_\mr{B}/7)}\right]^{1/2} \dot{M}_\mr{B}\,.
\label{eq:4}
\end{align}
Here, we take $R_\mr{in} \approx r_\mr{B}/7$ of our simulation setup
(see Table~\ref{tab:model}) as a reference value.

The mass supply rate to the accretion disc can be conservatively
estimated by $\dot{M} - \dot{M}_\mr{loss}(<R_\mr{in})$, meaning that
$\dot{M}$ measured in the simulation slightly overestimates the true
value.  This would be alleviated by taking a smaller value for
$R_\mr{in}$.  We have performed a test run with the smaller sink radius
(see Appendix~\ref{sec:res_check}), but found no remarkable differences
of the accretion rate.  Note that the effect of the angular momentum
becomes important on the smaller scale. It is not allowed to take an
arbitrary small sink radius without considering the effect of the
angular momentum.

If the accretion disc is spatially resolved, we expect further mass loss
happens due to, e.g., the disc winds
\citep[e.g.,][]{Blandford:1999aa,Zahra-Zeraatgari:2016aa,Begelman:2016aa}
and/or jets from a close vicinity of the BH
\citep[e.g.,][]{Ohsuga:2005aa,Jiang:2014aa,Yuan:2015aa,Sc-adowski:2016aa}.
Outflows from the sink caused by such phenomena may change the outer gas
dynamics on the scale of the Bondi radius. This should be studied in
future work.

\subsection{Parameter dependence}
\label{sec:dependence}

Here, we study how the flow structure changes with variation of the
simulation parameters: the shadow opening angle $\theta_\mr{shadow}$ (in
Sec.~\ref{sec:sdep}), BH mass $M_\mr{BH}$ (in Sec.~\ref{sec:Mdep}), and
ambient density $n_\infty$ (in Sec.~\ref{sec:ndep}).  In
Sec.~\ref{sec:comp_prev}, we compare our results with previous 1D
calculations.

\subsubsection{Dependence on shadow size}
\label{sec:sdep}

  \begin{table}
   \centering
   \caption{Summary of the $\theta_\mr{shadow}$ dependence.}
   \label{tab:s-model}
   \begin{tabular}{lccc} \hline
run&
$\theta_\mr{shadow}$$^{a}$& 
$\theta_\mr{inflow}(r_\mr{B})$$^{b}$ &
$\dot{M}/\dot{M}_\mr{B}$$^{c}$
 \\\hline
s100 (Dds)   & $45^\circ$    & $40^\circ$ & $59\%$\\
s075 & $37.75^\circ$ &$29^\circ$  & $42\%$\\
s050  & $25^\circ$    &$19^\circ$  & $25\%$\\
s025 & $11.25^\circ$ &$9^\circ$   & $6.5\%$\\\hline
   \end{tabular}\\
\begin{flushleft}
NOTES.\textemdash 
$^{a}$shadow opening angle of our subgrid model (equation~\ref{eq:10});
$^{b}$opening angle of equatorial neutral inflow region at $r_\mr{B}$;
$^{c}$accretion rate normalized by Bondi one.
\end{flushleft}
  \end{table}

\begin{figure}
\centering \includegraphics[trim=0 0 0 0, width=8.5cm,
clip]{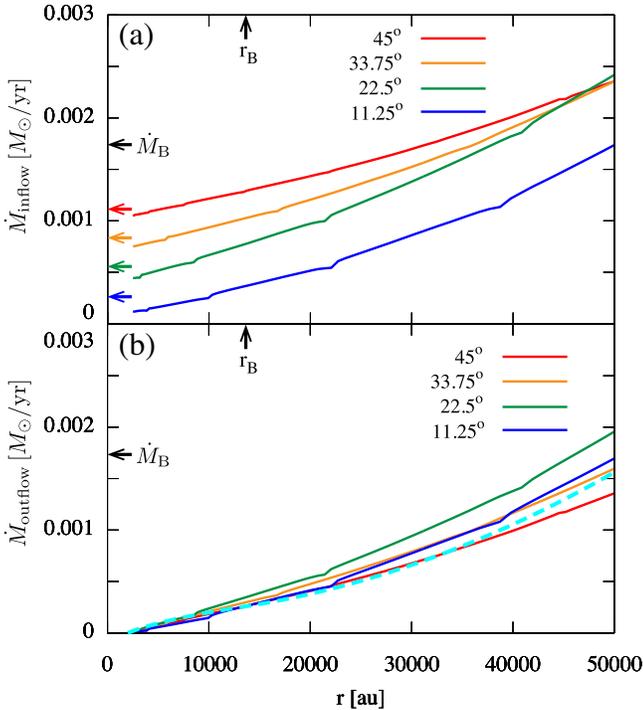} \caption{ Same as Fig.~\ref{fig:mdoio_Ds}
but for the runs s100 (same as Dds; red), s075 (orange), s050 (green)
and s025 (blue), with the shadow opening angles
$\theta_\mr{shadow}=45^\circ$, $33.75^\circ$, $22.5^\circ$ and
$11.25^\circ$, respectively.  (a) The inflow rate
$\dot{M}_\mr{inflow}(r)$ and (b) the outflow rate
$\dot{M}_\mr{outflow}(r)$ are plotted.  The horizontal arrows in panel
(a) mark the analytical estimates of $\dot{M}_\mr{inflow}(R_\mr{in})$
given by equation~\eqref{eq:17}.  The dashed cyan line in panel (b)
shows the analytical profile of $\dot{M}_\mr{outflow}(r)$ given by
equation~\eqref{eq:14}.}  \label{fig:mdotio_sdep}
\end{figure}

Considering uncertainties in the anisotropic shadowing effect
(Sec.~\ref{sec:direction_dep}), we study the cases with different shadow
opening angles $\theta_\mr{shadow}$ by reducing it from $45^\circ$ in
``Dds run'' (here we also call it ``s100 run'') to $33.75^\circ$ (``s075
run''), $22.5^\circ$ (``s050 run'') and $11.25^\circ$ (``s025 run'').
We call this series of runs as ``s-series''. We take
$M_\mr{BH}=10^3M_\odot$ and $n_\infty=10^5\cmr{cm^{-3}}$ for the
s-series.

Our main findings are as follows: in all the runs of the s-series, the
overall flow structures are similar and the accretion rates $\dot{M}$
are much higher than in the cases without the shadow (i.e., Di and Ddn
runs).  The obtained accretion rates and opening angles of the
equatorial neutral region at Bondi radius $\theta_\mr{inflow}(r_\mr{B})$
are summarized in Table~\ref{tab:s-model}.  The values of
$\theta_\mr{inflow}(r_\mr{B})$ agree well with the prediction by
equation~\eqref{eq:6} with only small offsets $\lesssim 3^\circ$.  Note
also that $\theta_\mr{inflow}(r_\mr{B}) \simeq \theta_\mr{shadow}$
despite the gradual transition between the shadowed and non-shadowed
regions modeled as in equation~\eqref{eq:10}.

The equatorial inflow rates $\dot{M}_\mr{inflow} (r)$ (see
equation~\ref{eq:16}) are shown in Fig.~\ref{fig:mdotio_sdep}(a).  The
values of $\dot{M}_\mr{inflow}$ at $R_\mr{in}$ agree well with the rates
estimated by the Bondi flow through the solid angle of
$4\pi\sin\theta_\mr{inflow}(r_\mr{B})$ (equation~\ref{eq:17}; arrows in
Fig.~\ref{fig:mdotio_sdep} a), but with slight downward offset due to
the photoevaporation mass loss.  Fig.~\ref{fig:mdotio_sdep}(b) shows the
outflow rates in polar directions $\dot{M}_\mr{outflow}(r)$ (again, see
equation~\ref{eq:16}).  As seen in Sec.~\ref{sec:shadow_rad}, the
estimate by equation~\eqref{eq:14} with $f_\mr{outflow}=0.7$ gives a
good fit to the numerical results.  Small differences of $\sim$ a few
$\times$ 10~\% among them are comparable to the intrinsic fluctuations
of the outflow rates present in the quasi-steady states.

\begin{figure}
\centering \includegraphics[width=8.5cm]{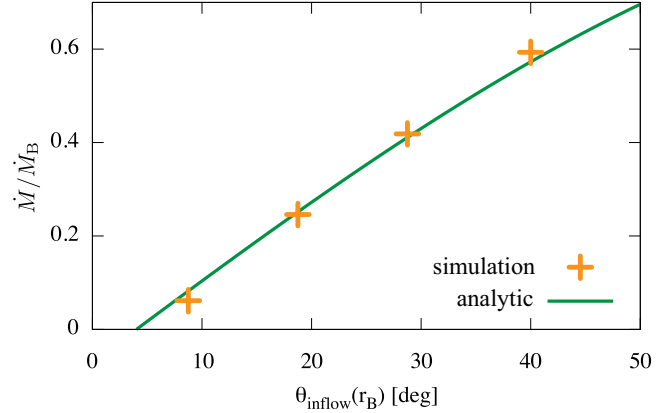}
 \caption{The net accretion rate $\dot{M}$ normalized by the Bondi rate
 $\dot{M}_\mr{B}$ as a function of the opening angle of the horizontal
 neutral layer at the Bondi radius $\theta_\mr{inflow}(r_\mr{B})$.  The
 crosses show the runs with shadow opening angles
 $\theta_\mr{shadow}=45^\circ$, $33.75^\circ$, $22.5^\circ$ and
 $11.25^\circ$. The solid line represents the relation given by
 equation~\eqref{eq:18} assuming $\dot{M}_\mr{loss}=0.07\dot{M}_\mr{B}$.
 }  \label{fig:mdot_th_indep}
\end{figure}

The net accretion rates $\dot{M}$ are plotted as crosses in
Fig.~\ref{fig:mdot_th_indep} against $\theta_\mr{inflow}(r_\mr{B})$.
Since equation~\eqref{eq:17} slightly overestimates
$\dot{M}_\mr{inflow}(R_\mr{in})$ due to the photoevaporation mass loss,
we modify equation~\eqref{eq:17} assuming a constant mass loss rate
$\dot{M}_\mr{loss}$ in all cases, as
\begin{align}
 \dot{M} = \dot{M}_\mr{inflow}(R_\mr{in}) =
\frac{\Delta\Omega_\mr{inflow}(r_\mr{B})}{4\pi}\dot{M}_\mr{B}
- \dot{M}_\mr{loss} \,.
\label{eq:18}
\end{align}
We find that $\dot{M}_\mr{loss}=0.07\,\dot{M}_\mr{B}$ gives the best fit
to the simulated results with errors less than 2\% of $\dot{M}_\mr{B}$.
This good agreement also supports the above assumption of constant
$\dot{M}_\mr{loss}$.  The value of $\dot{M}_\mr{loss}$ is similar to but
smaller than $\dot{M}_\mr{outflow}(r_\mr{B})$ (equation~\ref{eq:14})
partly due to the contribution from the circulation flows, as mentioned
in Sec.~\ref{sec:anl_modeling}.  Moreover, by setting $\dot{M} = 0$ in
equation~\eqref{eq:18}, we get the critical opening angle
$\theta_\mr{cr} \simeq 4^\circ$, below which the equatorial neutral flow
disappears by photoevaporation.  This value will be raised up to
$\theta_\mr{cr} \simeq 10^\circ$ if we include the mass loss inside the
sink, which is currently ignored (see Sec.~\ref{sec:mass_loss_inner}).

\subsubsection{Dependence on BH mass}
\label{sec:Mdep}

  \begin{table}
   \centering
   \caption{Summary of the $M_\mr{BH}$ dependence.}
   \label{tab:M-model}
   \begin{tabular}{lccc} \hline
run&
$M_\mr{BH}\,[M_\odot]$& 
$\theta_\mr{inflow}(r_\mr{B})$ &
$\dot{M}/\dot{M}_\mr{B}$
 \\\hline
M1e2 & $10^2$ & $38^\circ$ & $55\%$\\
M1e3 (Dds)   & $10^3$ & $40^\circ$ & $59\%$\\
M1e4 & $10^4$ & $43^\circ$ & $61\%$\\
M1e5 & $10^5$ & $48^\circ$ & $67\%$\\\hline
   \end{tabular}\\
  \end{table}

Next, we study the dependence on the BH mass $M_\mr{BH}$ by performing a
set of simulations termed ``M-series'', where $M_\mr{BH}=10^2\,M_\odot$
(``M1e2 run''), $10^3\,M_\odot$ (``M1e3 run'' identical to ``Dds run''),
$10^4\,M_\odot$ (``M1e4 run'' ) and $10^5\,M_\odot$ (``M1e5 run'').  The
other parameters are set to $n_\infty=10^5\cmr{cm^{-3}}$ and
$\theta_\mr{shadow}=45^\circ$.

We find that the accretion proceeds roughly at the Bondi rate in a
quasi-steady fashion for all the runs.  Flow properties at the end of
calculation are summarized in Table~\ref{tab:M-model}.  We see that, for
all the runs, the neutral region spans the opening angle
$\theta_\mr{inflow}(r_\mr{B})\simeq \theta_\mr{shadow} (=45^\circ)$ at
the Bondi radius, and that the Bondi-like accretion proceeds through
this solid angle with the rate $\dot{M} \simeq 0.5-0.7\,
\dot{M}_\mr{B}$.

Note, however, that the opening angle and thus the accretion rate
increase gradually with the BH mass.  This dependence can be understood
as follows. Recall that the luminosity $L$ is approximately proportional
to the Eddington value or the mass $M_\mr{BH}$ in the super-Eddington
regime (equations~\ref{eq:12} and \ref{eq:3}).  The radius of the \Hii
region thus varies as $r_\mr{HII}(\theta) \propto M_\mr{BH}^{2/3}$
(equation~\ref{eq:13}), while the Bondi radius follows $r_\mr{B} \propto
M_\mr{BH}$ (equation~\ref{eq:2}).  Since $r_\mr{HII}(\theta)$ is an
increasing function of $\theta$ (equation~\ref{eq:6}; see also
Fig.~\ref{fig:th_in_Ds}), this means that
$\theta_\mr{inflow}(r_\mr{B})$, obtained by solving
$r_\mr{B}=r_\mr{HII}(\theta)$ with respect to $\theta$, increases with
$M_\mr{BH}$.  Equation~\eqref{eq:6} indeed explains the variation of the
opening angle $\theta_\mr{inflow}(r_\mr{B})$ in the numerical results
within the error of $4^\circ$.  Similarly, the accretion rates
$\dot{M}/\dot{M}_\mr{B}$ estimated by equation~\eqref{eq:18} reproduce
the results with errors $\lesssim10\%$.

\subsubsection{Dependence on ambient density}
\label{sec:ndep}
  \begin{table}
   \centering
   \caption{Summary of the $n_\infty$ dependence.}
   \label{tab:n-model}
   \begin{tabular}{lccc} \hline
run&
$n_\infty\,[\mr{cm^{-3}}]$& 
$\theta_\mr{inflow}(r_\mr{B})$ &
$\dot{M}/\dot{M}_\mr{B}$
 \\\hline
n1e3 & $10^3$ & $36^\circ$ & $59\%$\\
n1e4 & $10^4$ & $38^\circ$ & $54\%$\\
n1e5 (Dds)   & $10^5$ & $40^\circ$ & $59\%$\\
n1e6 & $10^6$ & $44^\circ$ & $71\%$\\\hline
   \end{tabular}\\
  \end{table}

Motivated by a wide variety of the environment in the vicinity of BHs,
we finally investigate the cases with different ambient densities,
termed ``n-series'', where $n_\infty$ is $10^3\cmr{cm^{-3}}$ (``n1e3
run''), $10^4\cmr{cm^{-3}}$ (``n1e4 run''), $10^5\cmr{cm^{-3}}$ (``n1e5
run'' identical to ``Dds run'') and $10^6\cmr{cm^{-3}}$ (``n1e6 run'').
The other parameters are set to $M_\mr{BH}=10^3\,M_\odot$ and
$\theta_\mr{shadow}=45^\circ$.
  
The flow characteristics are similar regardless of $n_\infty$ with the
neutral-region opening angle $\theta_\mr{inflow}(r_\mr{B})\simeq
\theta_\mr{shadow} (=45^\circ)$ and the mass accretion rates comparable
to the Bondi rates, $\dot{M}/\dot{M}_\mr{B} \simeq 0.6-0.7$, in all the
cases (Table~\ref{tab:n-model}).  The increasing trend of
$\theta_\mr{inflow}(r_\mr{B})$ with $n_\infty$ can be understood again
as in Sec.~\ref{sec:Mdep}.  Now the luminosity $L$ is almost independent
of $n_\infty$ in the super-Eddington regime (equation~\ref{eq:3}), which
leads to $r_\mr{HII}(\theta) \propto n_\infty^{-2/3}$
(equation~\ref{eq:6}) while $r_\mr{B}$ is independent of $n_\infty$
(equation~\ref{eq:2}).  Since $r_\mr{HII}(\theta)$ is an increasing
function of $\theta$, it follows that $\theta_\mr{inflow}(r_\mr{B})$
increases with $n_\infty$.  Again, the analytic estimates of
$\theta_\mr{inflow}(r_\mr{B})$ with equation~\eqref{eq:6} and
$\dot{M}/\dot{M}_\mr{B}$ with equation~\eqref{eq:18} agree well with the
numerical results, with errors less than $4^\circ$ and $10\%$,
respectively.

\subsection{Comparison with previous works}
\label{sec:comp_prev}

Recently, \citet{Inayoshi:2016ac} and \citet{Sakurai:2016aa} have
investigated necessary conditions to overcome the radiative feedback to
achieve the Bondi-like accretion using 1D calculations under the
spherical symmetry.  These authors simulated accretion on to the BH in
the same setting but with different prescriptions on the BH irradiation,
i.e., whether or not the BH luminosity is capped by $L_\mr{E}$.  They
concluded that the efficient Bondi-like accretion appears when the
following condition is satisfied: $M_\mr{BH}\,n_\infty \gtrsim 10^9\,
M_\odot\,\mr{cm^{-3}}$ (see Sec.~\ref{sec:sph_acc}).

However, our 2D simulations suggest that the above criterion needs to be
modified. We find the efficient accretion at about the Bondi rate is
always possible, as long as the shadow opening angle has a certain size
($\theta_\mr{shadow} \gtrsim O(10)^\circ$), because the inflows from
equatorial shadowed regions are allowed in 2D simulations. We here
emphasize that neither $M_\mr{BH}$ nor $n_\infty$ appears in this
condition, and that the efficient accretion is possible even with
$M_\mr{BH}\,n_\infty \ll 10^9\, M_\odot\,\mr{cm^{-3}}$.  The accretion
rate through the shadowed direction is set by the Bondi rate, which
depends on $M_\mr{BH}$ and $n_\infty$.  For the accretion rate to
largely exceeds the Eddington rate, $\dot{M}_\mr{B}/\dot{M}_\mr{E}=
\left(M_\mr{BH}\,n_\infty/10^5\,M_\odot\,\mr{cm^{-3}}\right) \gg 1$
should be met in addition to the above condition for the shadow size.
Note, however, this condition for $M_\mr{BH}\,n_\infty$ is much easier
to be satisfied than the condition for the 1D calculations, i.e.,
$M_\mr{BH}\,n_\infty \gtrsim 10^9\, M_\odot\,\mr{cm^{-3}}$.

Our 2D simulations show that the flow structure qualitatively differs
from that in 1D even in the cases with $M_\mr{BH}\,n_\infty \gtrsim
10^9\, M_\odot\,\mr{cm^{-3}}$.  For example, with
$M_\mr{BH}=10^5\,M_\odot$ and $n_\infty=10^5\cmr{cm^{-3}}$ (M1e5 run),
the large bipolar \Hii bubbles persist in a steady state, whereas the
spherical Bondi-like accretion quenches the \Hii bubble in the 1D test
run.  This is partly because the enhanced ionizing radiation in the
polar directions due to the assumed directional dependence increases the
size of the \Hii bubbles.

As seen above, $M_\mr{BH}\,n_\infty$ is not the key parameter to
demarcate the regimes for efficient/inefficient accretion in 2D
simulations, unlike in the 1D cases.  The efficient accretion is
possible if only the shadow size is sufficiently large, irrespective of
$M_\mr{BH}$ or $n_\infty$.

\section{conclusions and discussion}
\label{sec:conclusion}

We have studied the black hole (BH) accretion of the primordial gas
under anisotropic irradiation by the circum-BH accretion disc.  Using
two-dimensional radiation hydrodynamics simulations, we have solved the
dynamics of the accretion flow spatially resolving both Bondi radius and
the size of the \Hii region, which can differ by 3-4 orders of
magnitude.  We do not resolve the central accretion disc which emits
anisotropic radiation, but inject ionizing photons at the inner boundary
of the computational domain according to the subgrid prescription.  To
see how the anisotropy of the BH irradiation affects the flow structure,
we first perform simulations with the three different types of the
directional dependence: isotropic radiation, anisotropic radiation from
the disc with and without the shadowing effect.  For the case with the
anisotropic shadowing effect, we have also studied the dependence of the
flow structure on the shadow opening angle $\theta_\mr{shadow}$, BH mass
$M_\mr{BH}$, and ambient density $n_\infty$.

With the isotropic irradiation, the accretion rate varies periodically
as a result of recurrent formation and collapse of a hot and low-density
\Hii bubble around the BH.  The time-averaged accretion rate is only
0.2\% of the original Bondi rate $\dot{M}_\mr{B}$ for the neutral medium
and roughly given by the Bondi rate for the ionized medium
\citep[e.g.,][]{Park:2011aa,Park:2012aa,Milosavljevic:2009ab}.  Even
with the anisotropy of the BH irradiation, the accretion rate is still
similar to that in the isotropic case unless the shadowing effect is
included.  The flow structure in this case, however, is qualitatively
different from the isotropic case: the large periodic variation, which
has been reported in previous studies, disappears.

Unlike in the former two cases, the accretion rate becomes much higher
in the case with the shadowing effect.  For example, in the case with
$M_\mr{BH}=10^3\,M_\odot$, $n_\infty=10^5\cmr{cm^{-3}}$ and
$\theta_\mr{shadow}=45^\circ$, the accretion rate reaches as high as
60\% of the Bondi rate $\dot{M}_\mr{B}$ and is ``super-critical'' with
400 times larger than the Eddington-limited rate $\dot{M}_\mr{E}$.  The
flow structure in the steady state consists of the equatorial Bondi-like
neutral inflow and bipolar ionized outflow.  Since the radiation is
confined to the polar directions, the rapid accretion proceeds in spite
of the BH luminosity eight times larger than the Eddington value.

We have investigated such steady flow structure with the analytical
models.  The opening angle of the equatorial neutral layer
$\theta_\mr{inflow}$ is derived from the balance between the supply and
consumption rates of ionizing photons in each direction
(equation~\ref{eq:6}).  In turn, the accretion rate $\dot{M}$ is modeled
assuming a Bondi-like flow through this equatorial layer also
considering the photoevaporation mass loss from its surfaces
(equation~\ref{eq:18}).  We have also found that, in order for the
equatorial Bondi-like inflow to be maintained, $\theta_\mr{inflow}$ at
the Bondi radius must be above a critical value, which is $\simeq
4^\circ$ for $M_\mr{BH}=10^3\,M_\odot$ and $n_\infty=10^5\cmr{cm^{-3}}$.
This value is raised up to $\sim 10^\circ$ if we account for the mass
loss inside the sink.  The parameter dependence of the flow structure
found in our simulations is well reproduced by this analytical model.

Our results highlight the importance of the directional dependence of BH
irradiation, especially in the equatorial directions, in determining
$\dot{M}$.  However, our current knowledge about the actual anisotropy
is very limited. Although not exactly the system of our interest,
line-driven disc winds around a supermassive BH (SMBH) with $M_\mr{BH}
\sim 10^8\,M_\odot$ is shown to create anisotropic radiation fields by
blocking high-energy photons
\citep[e.g.,][]{Proga:2000aa,Proga:2004aa,Nomura:2016aa}.
\citet{Proga:2000aa} suggest that the opening angle of the resulting
shadow is $\theta_\mr{shadow} \simeq 12^\circ$, which is larger than the
critical angle.  In reality, the accretion disc may undergo the
precession with variable angular momentum of accreting gas and change
the orientation of the shadowed region in time, resulting in the
destruction of the pre-existing neutral inflowing region.  In any case,
it is clearly awaited to study the structure of the inner part and
resulting anisotropy of the BH irradiation.  Our current study is
complementary to such future works, because our results provide outer
boundary conditions for them.

We have found that the required condition for the rapid accretion is
substantially relaxed from that obtained for the isotropic irradiation.
For these cases, the accretion rate is reduced to $\lesssim
0.01\,\dot{M}_\mr{B}$ by the radiative feedback unless the condition
$(M_\mr{BH}/10^4M_\odot)(n_\infty/10^5\cmr{cm^{-3}})\gtrsim 1$
\citep[e.g.,][]{Inayoshi:2016ac} is satisfied.  This condition requires
the ambient density $n_\infty$ as high as $10^6\cmr{cm^{-3}}$ even for
the most massive Pop III remnants with $M_\mr{BH}\sim 10^3M_\odot$ in
\cite{Hirano:2015aa}.  Such high ambient density seems difficult to
achieve because the typical central density of the first galaxies at
$z\sim 15$ is estimated as $10^5\cmr{cm^{-3}}$
\citep[e.g.,][]{Oh:2002aa,Volonteri:2005aa}, although it is
theoretically possible if the BH resides at the very centre of a halo
with the density profile $\rho \propto r^{-2}$
\citep[e.g.,][]{Wise:2007aa,Inayoshi:2016ac}.  With the shadowing
effect, BHs in a central part of the first galaxies can grow much more
quickly.

A long-term evolution of such fast mass growth would be as follows.
Suppose that a seed BH with mass $M_\mr{BH}=10^3M_\odot$ is embedded in
an ambient medium with $n_\infty= 10^5\cmr{cm^{-3}}$.  Using the shadow
opening angle of $\theta_\mr{shadow}=12^\circ$ suggested by
\citet{Proga:2000aa} for line-driven SMBH winds, we obtain the accretion
rate of $\dot{M} \sim 0.1\,\dot{M}_\mr{B} \sim
2\times10^{-4}(M_\mr{BH}/10^3M_\odot)^2\, M_\odot\cmr{yr^{-1}}$ with
equations~\eqref{eq:1} and \eqref{eq:18}.  Integrating this expression,
we obtain the growth history of the BH mass as
\begin{equation}
M_\mr{BH}(t)\sim  \frac{10^3\,M_\odot}{1-\left[(t-t_0)/5\cmr{Myr}\right]}\,,
\end{equation} 
where $t_0$ is the initial time of the accretion.  At a face value, the
BH mass diverges within a short timescale of $5\cmr{Myr}$.  In reality,
however, the BH mass growth via accretion should be limited by changes
in the environmental conditions, such as exhaustion of the ambient gas
by accretion.  If the remnant BHs of Pop III stars with $\sim
10^{2-3}\,M_\odot$ grow immediately in a few Myr time-scale by accretion
to $\sim 10^{5-6} M_\odot$, they subsequently evolve in the same way as
direct collapse BHs and can eventually grow to $\sim 10^9M_\odot$ SMBHs
via gas accretion and/or mergers by $z \sim 7$ \citep[see, e.g.,
][]{Tanaka:2009aa}.  With the shadowing effect, Pop III remnants can be
seeds for high-$z$ SMBHs.  It does not mean, however, that all the Pop
III remnants experience such rapid growth.  For example, if they stay in
a low-density region with $n<10 \cmr{cm^{-3}}$, as suggested by
\cite{Alvarez:2009aa}, they hardly grow in mass even at the Bondi
accretion rate.

Although we have assumed the weak rotation of the ambient gas, with
which a BH accretion disk should be much smaller than the Bondi radius,
accreting gas may have higher amount of the angular momentum in general.
Regarding the SMBH accretion, \cite{Li:2013aa} has shown that the
accretion rate is considerably reduced by the rotation \citep[see
also][]{Proga:2003ab,Proga:2003aa}.  If the flow predominantly comes
from the equatorial plane, the amount of the angular momentum carried,
and hence the impact of the rotational support, would be increased. To
investigate this effect, mechanisms for the angular momentum transfer
should also be considered (see below).

Our simulations also neglect the gas self-gravity.  The torque caused by
the self-gravity can play an important role in the angular momentum
transport \citep[see, e.g.,][]{Shlosman:1989aa,Shlosman:2016aa}. In
addition, the inward force of the self-gravity can enhance the gas
accretion on to BHs \citep[e.g.,][]{Li:2011aa}, on the scales larger
than both of the following two: (1) the radius where the enclosed gas
mass equals to the BH mass, $r_\mr{eq}=(3\,M_\mr{BH}/4\pi\rho)^{1/3}\sim
10^5\, (M_\mr{BH}/10^3\,
M_\odot)^{1/3}(n/10^5\,\cmr{cm^{-3}})^{-1/3}\,\mr{AU}$, and (2) the
Jeans length $\lambda_\mr{J}=\sqrt{\pi}c_\mr{s}/\sqrt{\rho G}\sim
10^6\,(n/10^5\,\cmr{cm^{-3}})^{-1/2}\,\mr{AU}$.  In our cases, however,
the Bondi radii are smaller than those scales and thus the flow
structures on the scale of Bondi radius are hardly affected by this
effect, although the size of \Hii bubbles can exceed them in some cases.

To be realistic, the following improvements are needed.  First of all,
the gas dynamics should be followed in 3D, in particular, to see the
effect of the gravitational torque.  Next, diffuse recombination photons
can modify the structure of neutral layer.  Furthermore, solving the gas
dynamics beyond our computational domain, i.e, in the outer molecular
and photodissociation regions \citep[e.g.,][]{Ricotti:2001aa}, will be
needed to see the large-scale flow structure.  Finally, it is crucial to
perform numerical simulations for the inner part dedicated to resolving
the generation of anisotropic radiation fields, which are inevitably
coupled to our simulations through the outer boundary conditions for
them.  We have clearly shown that the interplay of multi-scale processes
is essential in understanding the BH accretion.

\section*{Acknowledgements}

The authors would like to thank Kazumi Kashiyama, Rohta Takahashi,
Sanemichi Takahashi and Kenji Toma for fruitful discussions.  The
numerical simulations were performed on the Cray XC30 at CfCA of the
National Astronomical Observatory of Japan, as well as on the computer
cluster, {\tt Draco}, at Frontier Research Institute for
Interdisciplinary Sciences of Tohoku University.  This work is supported
in part by MEXT/JSPS KAKENHI Grant Number 15J03873 (KS), 25800102,
15H00776 and 16H05996 (TH), 15H06022 (HY) and 25287040 (KO).



\appendix
\section{details of chemical and thermal modelling}
\label{sec:chem_detail}
\subsection{Reaction rates}
\label{sec:reaction_rates}
  \begin{table*}
   \centering
   \caption{Chemical Reactions}
   \label{tab:reaction_rates}
   \begin{tabular}{lllc} \hline
No. & Reaction & Rate coeff. $\mr{[cm^3\,s^{-1}]}$& Ref. \\ \hline
1 & $\mr{H}+\mr{e} \rightarrow \mr{H^+} + 2\mr{e}$ &
$k_1=$
\parbox[t]{10cm}{
$\exp\left[-32.71396786 + 13.536556 \ln T_\mr{eV} -5.73932875 (\ln T_\mr{eV})^2\right.$\\
$+ 1.56315498 (\ln T_\mr{eV})^3 - 0.2877056 (\ln T_\mr{eV})^4 $\\
$+ 3.48255977\times 10^{-2} (\ln T_\mr{eV})^5-2.63197617\times 10^{-3} (\ln T_\mr{eV})^6 $\\
$\left.+ 1.11954395\times 10^{-4} (\ln T_\mr{eV})^7- 2.03914985\times 10^{-6} (\ln T_\mr{eV})^8\right]$}
 & 1\\ 
2 & $\mr{He}+\mr{e} \rightarrow \mr{He^+} + 2\mr{e}$ & 
$k_2=$
\parbox[t]{10cm}{
$\exp\left[-44.09864886 + 23.91596563\ln T_\mr{eV} - 10.7532302(\ln T_\mr{eV})^{2}\right.$\\
$+ 3.05803875(\ln T_\mr{eV})^{3} - 0.56851189(\ln T_\mr{eV})^{4}$\\
$+ 6.79539123\times10^{-2}(\ln T_\mr{eV})^{5} - 5.0090561\times10^{-3}(\ln T_\mr{eV})^{6}$\\
$\left. + 2.06723616\times10^{-4}(\ln T_\mr{eV})^{7} - 3.64916141\times10^{-6}(\ln T_\mr{eV})^{8}\right]$
}
& 1\\
3 & $\mr{He^+}+\mr{e} \rightarrow \mr{He^{2+}} + 2\mr{e}$ & 
$k_3=$
\parbox[t]{10cm}{
$\exp\left[-68.71040990 + 43.93347633*\ln T_\mr{eV} - 18.4806699(\ln T_\mr{eV})^{2}\right.$\\
$+ 4.70162649(\ln T_\mr{eV})^{3} - 0.76924663(\ln T_\mr{eV})^{4}$\\
$+ 8.113042\times10^{-2}(\ln T_\mr{eV})^{5} - 5.32402063\times10^{-3}(\ln T_\mr{eV})^{6}$\\
$\left.+ 1.97570531\times10^{-4}(\ln T_\mr{eV})^{7} - 3.16558106\times10^{-6}(\ln T_\mr{eV})^{8}\right]$
}
& 2\\
$4^{a}$ & $\mr{H^+}+\mr{e} \rightarrow \mr{H} + h\nu$ & 
$k_4=2.753\times10^{-14}  (T_\mr{K}/315614)^{-3/2}  (1+(T_\mr{K}/115188)^{-0.407})^{-2.242}$
& 3\\
$5^{b}$ & $\mr{He^+}+\mr{e} \rightarrow \mr{He} + h\nu$ & 
$k_5=k_\mr{5rr}+k_\mr{5di}$&\\
&&$k_\mr{5rr}=$
\parbox[t]{10cm}{
$T_\mr{K}^{-1/2}\exp\Big[\ln 10\times\big(
-10.47 - 0.1885\log_{10} T_\mr{K} $ \\
$+ 3.769\times10^{-2}(\log_{10} T_\mr{K})^2 
- 9.110\times10^{-3}(\log_{10} T_\mr{K})^3\big)\Big]$}
& 4\\
&&$k_\mr{5di}=1.9\times10^{-3}  T_\mr{K}^{-3/2} \exp[-473421/T_\mr{K}]  (1 + 0.3\exp[-94684/T_\mr{K}])$
& 5\\
$6^{c}$ & $\mr{He^{2+}}+\mr{e} \rightarrow \mr{He^+} + h\nu$ & 
$k_6=5.08\times10^{-13} (T_\mr{K}/40000)^{-0.8163-0.0208\log_{10}(T_\mr{K}/40000)}$
& 6\\
7 & $\mr{He^+}+\mr{H} \rightarrow \mr{He} + \mr{H^+}$ & 
$k_7=1.25\times10^{-15}(T_\mr{K}/300)^{0.25}$
& 7\\
8 & $\mr{H^+}+\mr{He} \rightarrow \mr{H} + \mr{He^+}$ & 
$k_8=$
\parbox[t]{10cm}{
\parbox[t]{7cm}{
$1.26\times10^{-9}T_\mr{K}^{-0.75}\exp[-127500/T_\mr{K}]$ }
$T_\mr{K}<10000$\\
\parbox[t]{7cm}{
$4.0\times10^{-37}T_\mr{K}^{4.74}$}
 $T_\mr{K}>10000$
}
& 8\\
9 & $2\mr{H} \rightarrow \mr{H^+} + \mr{H} + \mr{e}$ & 
$k_9=1.7\times10^{-4}k_1$
& 9\\
\hline
\end{tabular}\\
\begin{flushleft}
NOTES.\textemdash 
The $T_\mr{K}$ and $T_\mr{eV}$ are the gas temperature in units of K and eV, respectively;
$^{a}$Case B;
$^{b}$radiative \citep[Case B; singlet; our fit to][]{Hummer:1998aa} and dielectric \citep{Aldrovandi:1973aa} recombination;
$^{c}$Case B \citep[][with typo about the charge dependence corrected]{Draine:2011aa}.
\\
REFERENCES.\textemdash 
(1) \cite{Janev:1987aa};
(2) \citep[][from Aladdin database 1989]{Abel:1997aa};
(3) \cite{Ferland:1992aa};
(4) \cite{Hummer:1998aa};
(5) \cite{Aldrovandi:1973aa};
(6) \cite{Draine:2011aa};
(7) \cite{Zygelman:1989aa};
(8) \cite{Kimura:1993aa};
(9) \cite{Palla:1983aa}.
\end{flushleft}
\end{table*}

In Table~\ref{tab:reaction_rates}, we summarize the chemical reactions
considered in this work, which are adopted following
\cite{Glover:2007aa}, \cite{Abel:1997aa} and \cite{Anninos:1997aa}. We
adopt the Case B recombination rates for the recombination of
$\mr{H^+}$, $\mr{He^+}$ and $\mr{He^{2+}}$.  We neglect the $\mr{He^+}$
recombination through the quasi-stable triplet state $2^3S$ of
$\mr{He}$, assuming that $\mr{He}$ in that state is easily photoionized
by the BH irradiation \citep[see, e.g.,][]{Clegg:1989aa}.

\subsection{Cross sections}
\label{sec:cross_sections}
  \begin{table*}
   \centering
   \caption{Cross Sections}
\label{tab:cross_sections}
   \begin{tabular}{llllc} \hline
No. & Reaction & Cross section $\mr{[cm^2]}$ & & Ref. \\ \hline
1 & $\mr{H}+h\nu \rightarrow \mr{H^+} + \mr{e}$ & 
$\sigma_{\nu,1}=$
\parbox[t]{8cm}{
$6.30\times10^{-18}(\nu/\nu_\mr{T,1})^{-4}$\\
$\displaystyle\times\exp\left[4-
\frac{4\arctan((\nu/\nu_\mr{T,1}-1)^{1/2})}
{\displaystyle(\nu/\nu_\mr{T,1}-1)^{1/2}(1-\exp[-2\pi/(\nu/\nu_\mr{T,1}-1)^{1/2}])}\right]$}
&$h\nu_\mr{T,1}= 13.60\cmr{eV}$
& 1\\
2 & $\mr{He}+h\nu \rightarrow \mr{He^+} + \mr{e}$ & 
$\sigma_{\nu,2}=$
\parbox[t]{8cm}{
$3.14151\times10^{-16}
(\nu/\nu_\mr{T,2})^{7/2}
\Big(1 - 4.7416(\nu/\nu_\mr{T,2})^{-1/2}$\\
$ + 14.82(\nu/\nu_\mr{T,2})^{-1}- 30.8678(\nu/\nu_\mr{T,2})^{-3/2}+ 37.3584(\nu/\nu_\mr{T,2})^{-2}$\\
$
 - 23.4585(\nu/\nu_\mr{T,2})^{-5/2} + 5.9133(\nu/\nu_\mr{T,2})^{-3}\Big)$}
&$h\nu_\mr{T,2}= 24.58\cmr{eV}$
& 2\\
3 & $\mr{He^+}+h\nu \rightarrow \mr{He^{2+}} + \mr{e}$ & 
$\sigma_{\nu,3}=$
\parbox[t]{8cm}{
$1.575 \times10^{-18}(\nu/\nu_\mr{T,3})^{-4}$\\
$\displaystyle\times\exp\left[4-
\frac{4\arctan((\nu/\nu_\mr{T,3}-1)^{1/2})}
{\displaystyle(\nu/\nu_\mr{T,3}-1)^{1/2}(1-\exp[-2\pi/(\nu/\nu_\mr{T,3}-1)^{1/2}])}\right]$}
&$h\nu_\mr{T,3}= 54.40\cmr{eV}$

& 1\\ \hline
\end{tabular}\\
\begin{flushleft}
REFERENCES.\textemdash 
(1) \cite{Osterbrock:1989aa};
(2) \cite{Yan:1998aa}.
\end{flushleft}
\end{table*}

In Table~\ref{tab:cross_sections}, we summarize the cross sections
considered in this work.

\subsection{Heating and cooling rates}
\label{sec:heat_cool}
  \begin{table*}
   \centering
   \caption{Heating and Cooling Processes}
\label{tab:cool_rates}
   \begin{tabular}{lllc} \hline
No. & Process & Rate [$\mr{erg\, cm^{-3}\, s^{-1}}$] & Ref. \\ \hline
\multicolumn{4}{c}{Heating}\\ \hline
1 & $\mr{H}$ photoionization & $\Gamma_1$ (see text)&\\
2 & $\mr{He}$  photoionization & $\Gamma_2$ (see text)&\\
3 & $\mr{He^+}$ photoionization & $\Gamma_3$ (see text)&\\ \hline
\multicolumn{4}{c}{Cooling}\\ \hline
$1^{a}$ & $\mr{H^+}$ recombination & 
$\Lambda_1=$
\parbox[t]{10cm}{
$\exp\left[\ln 10\times\left(
-25.87 + 0.4958\log_{10} T_\mr{K} - 0.1052(\log_{10} T_\mr{K})^2 
\right.\right.
$\\
$
+4.264\times10^{-2}(\log_{10} T_\mr{K})^3 - 9.165\times10^{-3}(\log_{10} T_\mr{K})^4 
$\\
$
\left.\left.
+ 5.491\times10^{-4}(\log_{10} T_\mr{K})^5
\right)\right]\,n(\mr{e})\,n(\mr{H^+})$
}
& 1\\
$2^{b}$ & $\mr{He^+}$  recombination & 
\parbox[t]{11cm}{
$\Lambda_{2}=\Lambda_\mr{2rr}+\Lambda_\mr{2di}$\\
$\Lambda_\mr{2rr}=$
\parbox[t]{10cm}{
$T_\mr{K}^{1/2}\,\exp\left[\ln 10\times\left(
-26.22 + 0.4085\log_{10} T_\mr{K} + 0.1460(\log_{10} T_\mr{K})^2
\right.\right.
$\\
$\left.\left.
- 3.374\times10^{-2}(\log_{10} T_\mr{K})^3 + 1.733\times10^{-3}(\log_{10} T_\mr{K})^4
\right)\right]\,n(\mr{e})\,n(\mr{He^+})$
}
\\
$\Lambda_\mr{2di}=
1.24\times10^{-13}  T_\mr{K}^{-3/2}  (1 + 0.3\exp[-94000/T_\mr{K}]) \exp[-470000/T_\mr{K}]
\,n(\mr{e})\,n(\mr{He^+})
$
}
& 2,3\\
$3^{c}$ & $\mr{He^{2+}}$ recombination & 
$\Lambda_3= 1.38\times10^{-16}T_\mr{K} \left(0.684-0.0416\,\ln(T_\mr{K}/40000)\right)  
k_6\,n(\mr{e})\,n(\mr{He^{2+}})$
& 4 \\
4 & $\mr{H}$ excitation & 
$\Lambda_4=7.50\times10^{-19}  \left(1 + (T_\mr{K}/100000)^{1/2}\right)^{-1}  \exp[-118348/T_\mr{K}]
\,n(\mr{e})\,n(\mr{H})$
& 5\\
$5^{d}$ & $\mr{He}$  excitation & 
$\Lambda_5=1.1\times10^{-19}  T_\mr{K}^{0.082}  \exp[-230000/T_\mr{K}]
\,n(\mr{e})\,n(\mr{He})$
& 6\\
6 & $\mr{He^+}$ excitation & 
$\Lambda_6=5.54\times10^{-17} T_\mr{K}^{-0.397}  \left(1 + (T_\mr{K}/100000)^{1/2}\right)^{-1}  \exp[-473638/T_\mr{K}]
\,n(\mr{e})\,n(\mr{He^{+}})$
& 5\\
7 & $\mr{H}$ ionization & 
$\Lambda_7=2.18\times10^{-11}k_1\,n(\mr{e})\,n(\mr{H})$
& 7\\
$8^{c}$ & $\mr{He}$  ionization & 
$\Lambda_8=3.94\times10^{-11}k_2\,n(\mr{e})\,n(\mr{He})$
& 7\\
9 & $\mr{He^+}$ ionization & 
$\Lambda_9=8.72\times10^{-11}k_3\,n(\mr{e})\,n(\mr{He^{+}})$
& 7\\
10 &  Free-free &
\parbox[t]{11cm}{
\parbox[t]{10cm}{
$\Lambda_{10}=1.426\times10^{-27} T_\mr{K}^{1/2} 
\left(g_\mr{ff}(T_\mr{K};1) \left(n(\mr{H})+n(\mr{He^+})\right) + 4g_\mr{ff}(T_\mr{K};2)n(\mr{He^{2+}})\right)
\,n(\mr{e})$
}\\
$g_\mr{ff}(T_\mr{K};Z_\mr{i}) = $
\parbox[t]{10cm}{
$0.79464 + 0.1243  \log_{10}(T_\mr{K}/Z_\mr{i}^2)$\hspace{1cm}$T_\mr{K}/Z_\mr{i}^2<320000$\\
$2.13164 - 0.1240 \log_{10}(T_\mr{K}/Z_\mr{i}^2)$\hspace{1cm}$T_\mr{K}/Z_\mr{i}^2>320000$
}
}
 & 8\\
$11^{e}$ & Compton & 
$\Lambda_{11}=1.017\times10^{-37} T_\mr{CMB}^4  (T_\mr{K} - T_\mr{CMB}) \,n(\mr{e})$
& 5\\ \hline
\end{tabular}\\
\begin{flushleft}
NOTES.\textemdash 
$^{a}$Case B, our fit to \cite{Ferland:1992aa};
$^{b}$radiative \citep[Case B; singlet; our fit to][]{Hummer:1998aa} and dielectric 
\citep{Black:1981aa}
recombination cooling;
$^{c}$Case B \citep[][with typo about the charge dependence corrected]{Draine:2011aa};
$^{d}$singlet;
$^{e}$$T_\mr{CMB}=2.73(1 + z)$ with $z=15$.
\\
REFERENCES.\textemdash 
(1) \cite{Ferland:1992aa};
(2) \cite{Hummer:1998aa};
(3) \cite{Black:1981aa}
(4) \cite{Draine:2011aa};
(5) \cite{Cen:1992aa};
(6) \cite{Bray:2000aa};
(7) \cite{Anninos:1997aa};
(8) \cite{Shapiro:1987aa}.
\end{flushleft}
\end{table*}

In Table~\ref{tab:cool_rates}, we summarize the heating and cooling
processes considered in this work. Here, $n(X)$ is the number density of
species $X$ in units of $\mr{cm^{-3}}$.  We calculate the
photoionization heating rates as $\Gamma_i=\int (4\pi j_\nu/h\nu)
n(X_i)\sigma_{\nu,i} (h\nu - h\nu_{\mr{T},i})\mr{d}\nu$ (see
Table~\ref{tab:cross_sections}), with $X_1=\mr{H}$, $X_2= \mr{He}$ and
$X_3=\mr{He^{+}}$.

\section{Resolution check}
\label{sec:res_check}
\begin{figure}
\centering \includegraphics[width=8.5cm]{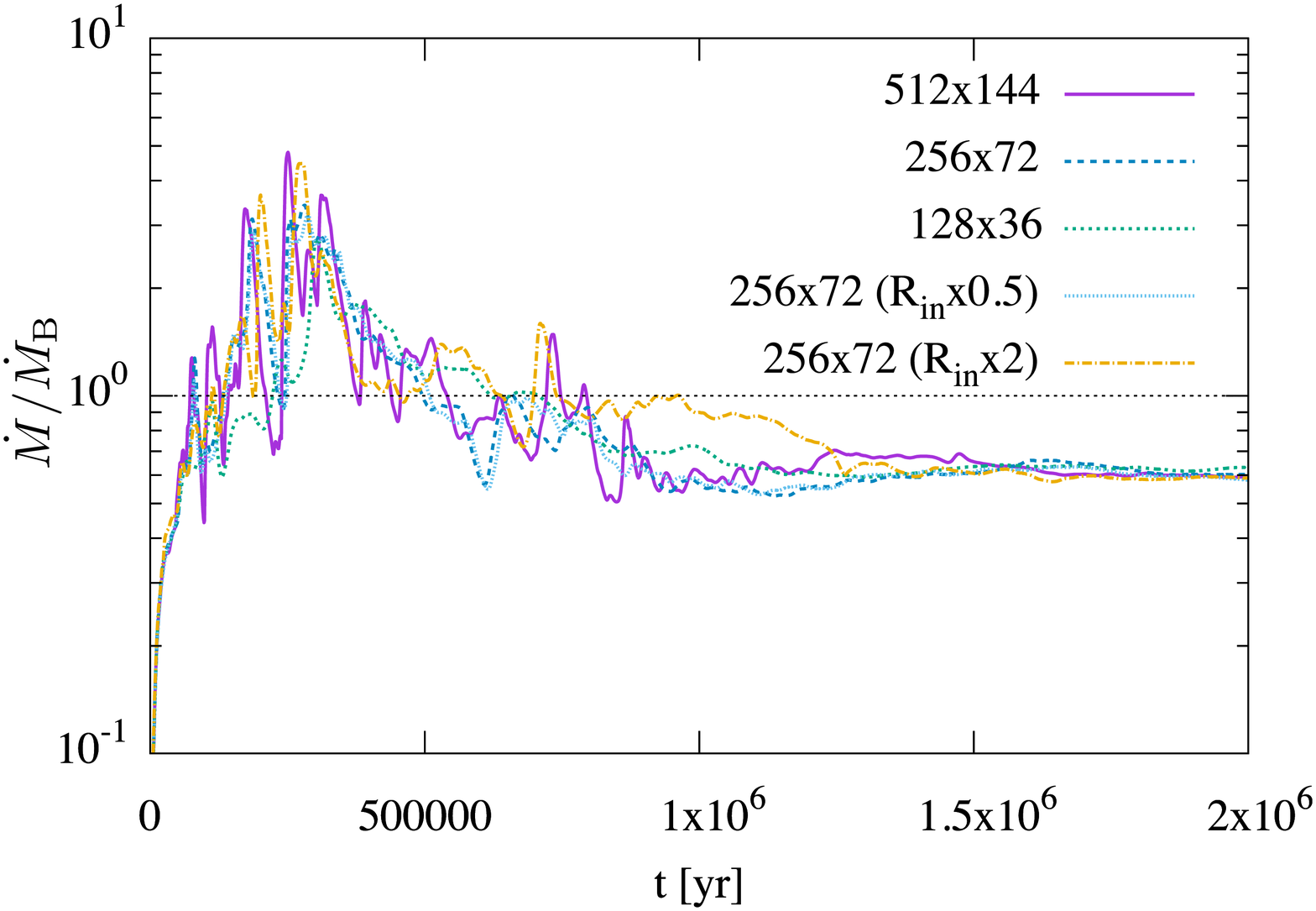}
 \caption{Same as Fig.~\ref{fig:mdot} but for the runs in
 App.~\ref{sec:res_check}.  The physical parameters are the same as Dds
 run but the resolution is different in each run.  See the text for
 details.} \label{fig:mdot_rdep}
\end{figure}

To check the resolution dependence of our results, we here see how the
evolution of $\dot{M}$ is affected by numerical settings, namely the
number of grids and sink size $R_\mr{in}$.  Taking the same physical
parameters as Dds run, we perform additional simulations with different
resolutions, as shown in Fig.~\ref{fig:mdot_rdep}.  Here, we take $N_r
\times N_\theta = 512\times 144$; $N_r \times N_\theta = 256\times 72$;
$N_r \times N_\theta = 128\times 36$; $N_r \times N_\theta = 256\times
72$ with $R_\mr{in}$ halved and doubled from the fiducial value.  Note
that our main results are obtained with the high- and medium-resolution
simulations with $N_r \times N_\theta = 512\times 144$ and $256\times
72$, respectively.

The dependence on the number of grids is checked by comparing the
results with $N_r \times N_\theta = 512\times 144$, $256\times 72$, and
$128\times 36$ (Fig.~\ref{fig:mdot_rdep}).  The strong variability of
$\dot{M}$ for $t \lesssim 10^6\cmr{yr}$ seen with the highest-resolution
is smoothed out with the lower resolutions. However, the values of
$\dot{M}$ at the end of the simulations are almost the same in all three
cases.  This confirms that the conclusion of this paper does not depend
on the number of grids.

Fig.~\ref{fig:mdot_rdep} also shows the evolution of $\dot{M}$ for the
cases with the different sink sizes.  The differences of accretion rates
are less than 10~\% for $t > 1.5 \times 10^6$ years in all three cases.
We also find that the values of $\dot{M}$ at the end of the simulations
decrease only by 4\% by halving the sink size from the fiducial value.
Such a trend is consistent with the estimated mass-loss rate from the
region between the halved and fiducial inner boundaries (see
equation~\ref{eq:14}), although it is also within the numerical error.
This ensures that the dependence of our results on the sink size is
weak.

\end{document}